\documentclass[twocolumn]{aastex631}

\usepackage{amsmath}
\graphicspath{{./}{figures/}}
\DeclareUnicodeCharacter{0308}{HERE!HERE!}

\shorttitle{SFR Enhancement in Interacting Galaxies}
\shortauthors{Shah et al.}

\received{July 19, 2022}
\revised{September 28, 2022}
\accepted{September 29, 2022}
\submitjournal{ApJ}

\begin{document}

\title{Investigating the Effect of Galaxy Interactions on Star Formation at \boldmath{$0.5<z<3.0$}}

\correspondingauthor{Ekta A. Shah}
\email{eas7266@rit.edu}

\author[0000-0001-7811-9042]{Ekta A. Shah}

\affil{Laboratory for Multiwavelength Astrophysics, School of Physics and Astronomy, Rochester Institute of Technology, 84 Lomb Memorial Drive, Rochester, NY 14623, USA}
\affil{Department of Physics and Astronomy, University of California,
Davis, One Shields Ave, Davis, CA 95616, USA}
\affil{LSSTC DSFP Fellow}

\author[0000-0001-9187-3605]{Jeyhan S. Kartaltepe}

\affil{Laboratory for Multiwavelength Astrophysics, School of Physics and Astronomy, Rochester Institute of Technology, 84 Lomb Memorial Drive, Rochester, NY 14623, USA}

\author[0000-0001-6333-8090]{Christina T. Magagnoli}
\affil{Laboratory for Multiwavelength Astrophysics, School of Physics and Astronomy, Rochester Institute of Technology, 84 Lomb Memorial Drive, Rochester, NY 14623, USA}

\author[0000-0002-1803-794X]{Isabella G. Cox}
\affil{Laboratory for Multiwavelength Astrophysics, School of Physics and Astronomy, Rochester Institute of Technology, 84 Lomb Memorial Drive, Rochester, NY 14623, USA}

\author{Caleb T. Wetherell}
\affil{Laboratory for Multiwavelength Astrophysics, School of Physics and Astronomy, Rochester Institute of Technology, 84 Lomb Memorial Drive, Rochester, NY 14623, USA}
 
\author[0000-0002-8163-0172]{Brittany N. Vanderhoof}
\affil{Laboratory for Multiwavelength Astrophysics, School of Physics and Astronomy, Rochester Institute of Technology, 84 Lomb Memorial Drive, Rochester, NY 14623, USA}

\author[0000-0002-2200-9845]{Kevin C. Cooke}
\affil{Laboratory for Multiwavelength Astrophysics, School of Physics and Astronomy, Rochester Institute of Technology, 84 Lomb Memorial Drive, Rochester, NY 14623, USA}
\affil{AAAS S\&T Policy Fellow hosted at the National Science Foundation,1200 New York Ave, NW, Washington, DC, US 20005}
\affil{Department of Physics \& Astronomy, University of Kansas, Lawrence, KS 66045, USA}

\author[0000-0003-2536-1614]{Antonello Calabro}
\affil{INAF OAR, via Frascati 33, Monte Porzio Catone 00078, Italy}

\author[0000-0003-3691-937X]{Nima Chartab}
\affil{Department of Physics and Astronomy, University of California, Riverside, 900 University Ave, Riverside, CA 92521, USA}

\author[0000-0003-1949-7638]{Christopher J. Conselice}
\affil{Centre for Particle Theory and Astronomy, University of Nottingham, Nottingham NG7 2RD, UK}

\author[0000-0002-5009-512X]{Darren J. Croton}
\affil{Centre for Astrophysics \& Supercomputing, Swinburne University of Technology, P.O. Box 218, Hawthorn, Victoria 3122, Australia}
\affil{ARC Centre of Excellence for All Sky Astrophysics in 3 Dimensions (ASTRO 3D)}

\author[0000-0002-6219-5558]{Alexander de la Vega}
\affil{Department of Physics and Astronomy, Johns Hopkins University, Baltimore, MD, 21218}

\author[0000-0001-6145-5090]{Nimish P. Hathi}
\affil{Space Telescope Science Institute, 3700 San Martin Dr., Baltimore, MD 21218, USA}

\author[0000-0002-7303-4397]{Olivier Ilbert}
\affil{Aix Marseille Universit\'e, CNRS, LAM (Laboratoire d’Astrophysique de Marseille) UMR 7326, 13388, Marseille, France}

\author[0000-0003-4268-0393]{Hanae Inami}
\affil{Hiroshima Astrophysical Science Center, Hiroshima University, 
1-3-1 Kagamiyama, Higashi-Hiroshima, Hiroshima 739-8526, Japan}

\author[0000-0002-8360-3880]{Dale D. Kocevski}
\affil{Department of Physics and Astronomy, Colby College, Waterville, ME 04961, USA}

\author[0000-0002-6610-2048]{Anton M. Koekemoer}
\affil{Space Telescope Science Institute, 3700 San Martin Dr., Baltimore, MD 21218, USA}

\author[0000-0002-1428-7036]{Brian C. Lemaux}
\affil{Department of Physics \& Astronomy, University of California, Davis, One Shields Ave., Davis, CA 95616, USA}
\affil{Gemini Observatory, 670 N. A'ohoku Place
Hilo, Hawaii, 96720, USA}

\author{Lori Lubin}
\affil{Department of Physics \& Astronomy, University of California, Davis, One Shields Ave., Davis, CA 95616, USA}

\author{Kameswara Bharadwaj Mantha}
\affil{Department of Physics and Astronomy, University of Missouri-Kansas City, Kansas City, MO 64110, USA}

\author[0000-0001-5544-0749]{Stefano Marchesi}
\affiliation{INAF - Osservatorio di Astrofisica e Scienza dello Spazio di Bologna, Via Piero Gobetti, 93/3, 40129, Bologna, Italy}
\affiliation{Department of Physics and Astronomy, Clemson University,  Kinard Lab of Physics, Clemson, SC 29634, USA}

\author[0000-0001-5454-1492]{Marie Martig}
\affil{Astrophysics Research Institute, Liverpool John Moores University, 146 Brownlow Hill, Liverpool L3 5RF, UK}

\author[0000-0002-3430-3232]{Jorge Moreno}
\affil{Department of Physics and Astronomy, Pomona College, 333 N. College Way, Claremont, CA 91711, USA}

\author[0000-0002-4140-0428]{Belen Alcalde Pampliega}
\affil{Departamento de Física de la Tierra y Astrofísica, Facultad de CC Físicas, Universidad Complutense de Madrid E-2840 Madrid, Spain}

\author[0000-0002-1871-4154]{David R. Patton}
\affil{Department of Physics and Astronomy, Trent University, 1600 West Bank Drive, Peterborough ON K9L 0G2, Canada}

\author[0000-0001-7116-9303]{Mara Salvato}
\affil{Max-Planck-Institut fu ̈r extraterrestrische Physik (MPE),
Giessenbachstrasse 1, D-85748 Garching bei Mu ̈nchen, Germany}

\author[0000-0001-7568-6412]{Ezequiel Treister}
\affil{Instituto de Astrofisica, Facultad de Fisica, Pontificia Universidad Catolica de Chile, Casilla 306, Santiago 22, Chile}

\begin{abstract}

Observations and simulations of interacting galaxies and mergers in the local universe have shown that interactions can significantly enhance the star formation rates (SFR) and fueling of Active Galactic Nuclei (AGN). However, at higher redshift, some simulations suggest that the level of star formation enhancement induced by interactions is lower due to the higher gas fractions and already increased SFRs in these galaxies. To test this, we measure the SFR enhancement in a total of 2351 (1327) massive ($M_*>10^{10}M_\odot$) major ($1<M_1/M_2<4$) spectroscopic galaxy pairs at $0.5<z<3.0$ with $\Delta V <5000$\thinspace km\thinspace s$^{-1}$ ($1000$\thinspace km\thinspace s$^{-1}$) and projected separation $<150$\thinspace kpc selected from the extensive spectroscopic coverage in the COSMOS and CANDELS fields.  We find that the highest level of SFR enhancement is a factor of 1.23$^{+0.08}_{-0.09}$ in the closest projected separation bin ($<25$\thinspace kpc) relative to a stellar mass-, redshift-, and environment-matched control sample of isolated galaxies. We find that the level of SFR enhancement is a factor of  $\sim1.5$ higher at $0.5<z<1$ than at  $1<z<3$ in the closest projected separation bin. Among a sample of visually identified mergers, we find an enhancement of a factor of 1.86$^{+0.29}_{-0.18}$ for coalesced systems. For this visually identified sample, we see a clear trend of increased SFR enhancement with decreasing projected separation (2.40$^{+0.62}_{-0.37}$ vs.\ 1.58$^{+0.29}_{-0.20}$  for $0.5<z<1.6$ and $1.6<z<3.0$, respectively). The SFR enhancement seen in our interactions and mergers are all lower than the level seen in local samples at the same separation, suggesting that the level of interaction-induced star formation evolves significantly over this time period.

\end{abstract}

\keywords{Galaxies: active, distances and redshifts, evolution, high-redshift, interactions, irregular}

\section{Introduction} \label{sec:obs_sfr_intro}

Galaxy interactions and mergers can have a substantial impact on the evolution of galaxies.  Simulations \citep[e.g.,][]{mihos1996,hopkins2008,dimatteo2008,scudder2012,moreno2015,moreno2019} of galaxies in the nearby universe show that interactions and mergers cause strong gravitational torques resulting in gas inflows towards the central regions of the galaxies, possibly resulting in nuclear starbursts and the triggering of AGN. This scenario is also supported by observations in the nearby universe. For example, most of the nearby Ultraluminous Infrared Galaxies (ULIRGs) and quasars ($>80\%$) show signatures of either an ongoing or recent galaxy merger \citep{sanders1988a,sanders1988b,urrutia2008}. Simulations also show an increased star formation rate (SFR) during an interaction throughout the galaxy, including in tidal tails caused by tidal interaction-induced accretion, redistribution, and compression of gas \citep{renaud2009,renaud2015,moreno2021}. Local galaxies with substantial tidal tails, such as the galaxy pair known as ``The Mice," are clear examples of this process \citep{barnes2004}.  

While galaxies that show strong visually identified morphological signatures of an interaction or merger can be used to study the late stages of galaxy interactions, these systems do not represent the complete merger sequence. This is because the resulting morphological signatures depend on many intrinsic factors of the interacting galaxies (e.g., their orbit, mass ratio, gas fraction, and initial morphology), and the observability of these signatures also depends on the redshift, viewing angle, depth, and wavelength of the observations \citep{lotz2011,blumenthal2020}. Hence, a spectroscopically confirmed kinematic galaxy pair sample identified based on close physical proximity of two galaxies (which may or may not show morphological merger signatures) is essential to develop a complete understanding of the merger process and its effect on galaxy properties.

Numerous studies in the local universe compare the SFR of interacting galaxies and mergers with isolated (control) galaxies \citep[e.g.,][]{larson1978,donzelli1997,bergvall2003,lambas2003,alonso2004,woods2007,ellison2008,knapen2009,robaina2009,darg2010,xu2010,ellison2013a,barrera-Ballesteros2015}.  Most of these studies find the largest increase in the SFRs of galaxies in pairs relative to their controls galaxies at projected separations of $<$30\thinspace kpc. They also find a trend of increasing relative SFRs with decreasing projected separations of pairs. Notably, \citet{patton2013} observe a significant level of SFR enhancement in local galaxy pairs with projected separations out to 150\thinspace kpc.

There is some evidence from simulations that galaxy interactions and mergers may not enhance SFRs to the same degree at higher redshift \citep[e.g.,][]{fensch2017,patton2020}. For example, based on an idealized binary simulation of galaxy mergers, \citet{fensch2017} find that the excess of merger-induced star formation and its duration are both about ten times lower at high redshift ($z\sim2$: gas fraction $\sim 60\%$, where gas fraction = $\frac{M_{gas}}{M_{gas}+M_*}$) compared to low redshift galaxies ($z\sim0$: gas fraction $\sim 10\%$). Other studies based on idealized binary simulations of mergers of galaxies with high gas fractions also suggest a lower peak and duration in SFR enhancement \citep{bournaud2011,hopkins2013,scudder2015}. \citet{perret2014} do not find any SFR enhancement in high redshift ($1<z<2$) galaxy mergers in the MIRAGE simulations. While \citet{patton2020} and \citet{martin2017} find some evidence for a decrease in SFR enhancement with increasing redshift ($0<z<1$) in interacting galaxies in the IllustrisTNG and Horizon-AGN cosmological simulations, respectively, \citet{hani2020} find no redshift dependence of SFR enhancement in post-merger galaxies ($0<z<1$) in IllustrisTNG. While many of these studies based on simulations find significant differences between interaction-induced SFR enhancement for high- and low-redshift interactions, these predictions have yet to be confirmed using through observations of high redshift galaxy pairs.

To date, there have been a few observational studies on star formation enhancement in high redshift mergers. \citet{kaviraj2013} find a sSFR enhancement of a factor of $\sim 2.2$ in major mergers in comparison with non-interacting galaxies at $z\sim2$. \citet{lackner2014} also estimate an enhancement factor of $\sim2$ in merging galaxies (projected separation between 2.2 and 8\thinspace kpc) compared to non-merging galaxies at $0.25<z<1.0$. However, using the same method, \citet{silva2018} identify merging galaxies (projected separation between 3 and 15\thinspace kpc) at $0.3<z<2.5$ and find no significant differences in the star formation of merging galaxies and non-merging galaxies. Using convolutional neural networks, \citet{pearson2019} identify more than 200,000 galaxies in the Sloan Digital Sky Survey (SDSS), the Kilo-Degree Survey (KiDS), and the CANDELS survey images as merging or non-merging and see a slight SFR enhancement factor of $\sim1.2$ in the merging galaxies over $0<z<4$. 

The change in the level of interaction-induced SFR enhancement with redshift becomes highly relevant when studying the role of galaxy mergers in galaxy evolution at cosmic noon ($1.5<z<3.0$),  i.e., around the peak epoch of the cosmic SFR density \citep{madau2014}. It is now well-established that most star-forming galaxies follow a SFR-M$_{*}$ relation often called the star-forming main sequence (SFMS) \citep{Brinchmann2004,elbaz2007,noeske2007,whitaker2014}. At low redshifts, the regime above the SFMS, i.e., the starburst galaxy population, is dominated by late-stage galaxy mergers \citep{sanders1988a,urrutia2008}. Though mergers are more frequent at cosmic noon \citep{Duncan2019}, some studies based on deep observations of high redshift galaxies suggest that mergers may not be the dominant cause of their increased star formation \citep[$z\sim2$; e.g.,][]{Rodighiero2011,lackner2014}. \citet{kaviraj2013} estimate that major mergers only contribute $\sim15\%$ to the overall SF budget, while \citet{osborne2020} and \citet{lofthouse2017} estimate even smaller values of SF contribution ($3-5\%$) from mergers. This lower contribution from mergers could be due to the fact that galaxies at high redshifts already have high SFRs, which could make it difficult to increase the SFR even further through mergers. The precise impact that galaxy interactions and mergers have on star formation at high redshift is still under debate.

The star formation process is directly tied to the properties of the gas in galaxies. The average gas fraction in galaxies changes substantially with redshift \citep[0.2-10\% at $z\sim0$, 40-60\% at $z$ $\sim 2$ in spiral galaxies;][]{daddi2010,tacconi2010,scoville2014}. High redshift galaxies also typically have a clumpier gas distribution with a higher average velocity dispersion ($\sim4\times$) than low redshift galaxies \citep{stott2016}. While having a much larger gas supply at high redshift could be useful for forming new stars and generating strong gas inflows, the high turbulence could make further compression of the gas and generation and propagation of inflows weaker than in low redshift interactions \citep[e.g.,][]{daddi2010,fensch2017}. Hence, there are multiple redshift-dependent factors that can affect the level of interaction-induced star formation enhancement, and studies of large samples of galaxies at high redshift are needed to determine the combined effect of these redshift-dependent factors on the interaction-induced SFR enhancement level of galaxies.

Here, we study the SFR enhancement in the largest sample of massive high-redshift spectroscopically confirmed major galaxy pairs (2351 pairs; \citealt{shah2020}) to date. We generated this sample using deep multiwavelength photometric and dedicated spectroscopic observations in the Cosmic Assembly Near-infrared Deep Extragalactic Legacy Survey  \citep[CANDELS;][]{grogin2011,koekemoer2011} and the Cosmic Evolution Survey \citep[COSMOS;][]{scoville2007} fields. We use a corresponding control sample generated by matching the stellar mass, redshift, and environment of isolated galaxies to individual paired galaxies as described in \citet{shah2020}.  We compare the SFRs of the two samples to estimate the interaction-induced SFR enhancement in paired galaxies. We also use the same method to study the star formation enhancement in visually identified samples of late-stage galaxy interactions and mergers selected from \citet{kartaltepemor15}.

The layout of this paper is as follows. We describe the observations and data products used for our analysis in Section \ref{sec:obs_sfr_data}. In Section \ref{sec:obs_sfr_sampleselection}, we describe the criteria for generating the pair sample, visually identified interaction and merger samples, and their corresponding control samples. We estimate the SFRs of galaxies in our sample in Section \ref{sec:sf-analysis} and compute the SFR enhancement and present our results in Section \ref{sec:sfr_enh}. We discuss our results in Section \ref{sec:obs_sfr_discussion} and summarize this study in Section \ref{sec:obs_sfr_summary}. Throughout this work, we use the standard $\Lambda$CDM cosmology with $H_0=70$\thinspace km s$^{-1}$ Mpc$^{-1}$, $\Omega_\Lambda=0.7$, and $\Omega_M= 0.3$. All magnitudes are given in the observed AB system and mass values of the galaxies correspond to their stellar masses unless stated otherwise.

\section{Data} \label{sec:obs_sfr_data}
  We use spectroscopic and multi-wavelength photometric observations from the CANDELS \citep[PIs: S. Faber and H. Ferguson;][]{grogin2011,koekemoer2011} and COSMOS \citep{scoville2007} surveys for our analysis. The deep and extensive datasets available in the fields observed by these surveys provide useful observations of a complete sample of massive galaxies with a stellar mass greater than 10$^{10}$\thinspace M$_{\odot}$ out to $z\sim3$. COSMOS is the largest contiguous area HST survey, covering a $\sim2$\,deg$^2$ area on the sky and  observing more than two million galaxies. Ancillary observations have now been obtained across the entire electromagnetic spectrum, covering over 40 photometric bands (described in more detail below). 
  
  The CANDELS survey consists of HST imaging of five different well-studied fields, including: (i) a portion of the COSMOS field, (ii) a portion of the UKIDSS Ultra-Deep Survey \citep[UDS;][]{lawrence2007}, (iii) the Great Observatories Origins Deep Survey \citep[GOODS;][]{giavalisco2004} -North (GOODS-N), (iv) GOODS-South (GOODS-S), and (v) the Extended Groth Strip \citep[EGS;][]{davis2007}. CANDELS observations provide F160W and F125W imaging taken with the HST/Wide Field Camera 3 (WFC3) and the F606W and F814W imaging with the HST/Advanced Camera for Surveys (ACS) as well as other ancillary observations in each of the five fields as described below.

\subsection{Photometric Observations}

We use CANDELS and COSMOS team-compiled photometric catalogs containing the positions, stellar masses, photometric redshifts, and fluxes of galaxies at different wavelengths  \citep[see details in][]{shah2020}.  The sources in both surveys were identified using the source detection algorithm \textsc{Source Extractor} \citep{bertin1996}.  The photometric catalogs for the full COSMOS field, the COSMOS-CANDELS region, UDS, GOODS-N, GOODS-S, and EGS are published in \cite{laigle2016}, \cite{nayyeri2017}, \cite{galametz2013}, \cite{guo2013}, \citet{barro2019}, and \cite{stefanon2017}, respectively. We use the near-UV--Far-Infrared (FIR) observations
to compute the SFR and stellar mass for galaxies in the large COSMOS \citep{laigle2016}, CANDELS-COSMOS \citep{sanders2007,ashby2013,nayyeri2017}, GOODS (N-S) \citep{dickinson2003,giavalisco2004,ashby2013}, EGS \citep{barmby2008,ashby2015} and UDS \citep{ashby2013,ashby2015} fields
using spectral energy distribution (SED) fitting as described in Section~\ref{subsec:sf-est}.

\subsubsection{Photometry and Physical Properties of Observed Galaxies}

We used the COSMOS2015 photometric catalog for the $\sim2$\,deg$^2$ COSMOS field, which includes photometry of more than half a million galaxies in over 30 bands as well as their estimated stellar mass and photometric redshift values \citep{laigle2016}. The photometric redshifts and stellar masses were estimated using the SED fitting code {\sc LePhare}\footnote{\url{http://www.cfht.hawaii.edu/~arnouts/LEPHARE/lephare.html}} \citep{arnouts2002,ilbert2006} to fit multiwavelength observations of galaxies. The fitting process is based on the \citet{chabrier2003} IMF, two metallicities (solar and half-solar), emission lines templates from \citet{ilbert2009},  attenuation curves from \citep{calzetti2000,arnouts2013}, an exponentially declining and delayed star formation history, and Stellar Population Synthesis models from \cite{bruzual2003}.

The CANDELS team combined the multiwavelength observations with different spatial resolutions and performed uniform photometry across different filters using TFIT \citep{laidler2007,lee2012} to generate the final photometric catalogs for each of the fields \citep{guo2013,nayyeri2017,galametz2013,stefanon2017,barro2019}. The team estimated the photometric redshifts of galaxies using the method of \citet{dahlen2013}, which combines the posterior probability distribution of photometric redshifts from various SED fitting codes and templates. The method chooses the best-estimated value of photometric redshift by selecting the median value of the peak redshifts of these different Probability Distribution Functions (PDFs). Similarly, ten different teams estimated the stellar mass of galaxies using different SED templates based on different galaxy populations \citep{santini2015,mobasher2015}. The median value of the average of all these PDFs was then selected as the best stellar mass estimate for a given galaxy.

The stellar mass measurements in the above photometric catalogs were based on the photometric redshifts of the galaxies.  We require spectroscopic redshifts for our sample spectroscopic galaxy pairs and controls, which could be different from the photometric redshift of a given galaxy. Therefore, we re-measure the stellar masses of our pairs and controls using the SED fitting code \citep[{\sc MAGPHYS};][]{dacunha2008} to fit the above-mentioned photometric observations of galaxies and their spectroscopic redshifts as described in detail in Section~\ref{subsec:sf-est}. For the galaxies that have similar photometric and spectroscopic redshifts, our newly estimated stellar masses are consistent with the original stellar masses from the photometric catalogs. We use these MAGPHYS estimated stellar masses to generate the galaxy pair sample and its corresponding control sample using the selection criteria described in Section \ref{sec:obs_sfr_sampleselection}. Since we use photometric redshifts to select the visually identified interactions, mergers, and their control samples, we use the original stellar mass measurements from the photometric catalogs rather than re-computing them.

\subsection{Spectroscopic Observations}

We used all the available existing spectroscopic observations (published, as shown in Table \ref{table:spec_obs}, and unpublished) compiled by the COSMOS and CANDELS teams to generate our spectroscopic galaxy pair sample and the corresponding control sample. We also used spectroscopic observations obtained using Keck II/DEIMOS \citep{shah2020}, Gemini/GMOS (I. Cox et al., in preparation), and Keck I/MOSFIRE (B. Vanderhoof et al., in preparation) for the UDS, COSMOS, and GOODS-S fields. We only used reliable spectroscopic redshifts, i.e., those with a quality flag of two or greater \citep{shah2020}, to generate our pair and control samples.

\begin{deluxetable*}{cc}
\tablewidth{0pt}
\tablecaption{Spectroscopic Observations}
\tablehead{\colhead{\textbf{Telescope/Instrument}} & \colhead{\textbf{Reference}}}
\startdata
\textbf{COSMOS}\\
\hline
VLT/VIMOS & \citet{lilly2007, tasca2015, lefevre2015}\\  
& \citet{vanderWel2016, straatman2018}\\
VLT/FORS2 & \citet{comparat2015, pentericci2018}\\
Keck I/MOSFIRE and LRIS & \citet{kriek2015, masters2019}\\
Keck II/DEIMOS & \citet{capak2004, kartaltepe2010}\\
 & \citet{hasinger2018, masters2019}\\
MMT/Hectospec spectrograph & \citet{damjanov2018}\\
Subaru/MOIRCS & \citet{onodera2012}\\
Subaru/FMOS& \citet{silverman2015, kartaltepe2015}\\
HST/WFC3-IR grism & \citet{krogager2014, momcheva2016}\\
Magellan (Baade) telescope/IMACS & \citet{trump2009, coil2011}\\
\hline
\textbf{UDS}\\
\hline
HST/WFC3-IR grism & \citet{morris2015, momcheva2016}\\ 
VLT/VIMOS and FORS2 & \citet{bradshaw2013, pentericci2018}\\ Keck I/MOSFIRE and LRIS & \citet{kriek2015, masters2019}\\ 
Keck II/DEIMOS & \citet{masters2019}\\
VLT/VIMOS & \citet{mclure2018, scodeggio2018}\\
\hline
\textbf{GOODS-N}\\
\hline
HST/WFC3-IR grism & \citet{ferreras2009, momcheva2016}\\ 
Keck I/MOSFIRE and LRIS & \citet{cowie2004, reddy2006}\\ 
 & \citet{barger2008, kriek2015, wirth2015}\\
Keck II/DEIMOS  & \citet{wirth2004, cowie2004, barger2008, cooper2011}\\
Subaru Telescope/MOIRCS & \citet{yoshikawa2010}\\
\hline
\textbf{GOODS-S}\\
\hline
VLT/VIMOS & \citet{lefevre2004, ravikumar2007, balestra2010}\\
 & \citet{lefevre2013, mclure2018}\\
VLT/FORS1 and FORS2 & \citet{daddi2004, szokoly2004, vanderwel2004}\\
 & \citet{mignoli2005, vanzella2008, popesso2009}\\ 
  & \citet{vanzella2008, vanzella2009, balestra2010}\\
   & \citet{kurk2013, pentericci2018}\\
 VLT/MUSE & \citet{inami2017, urrutia2019}\\ HST/WFC3-IR grism & \citet{ferreras2009, morris2015, momcheva2016}\\
Gemini/GMOS & \citet{roche2006}\\
Keck I/MOSFIRE & \citet{kriek2015}\\
Keck II/DEIMOS & \citet{silverman2010, cooper2012c}\\
AAT/LDSS++ spectrograph & \citet{croom2001}\\
\hline
\textbf{EGS}\\
\hline
 Keck I/MOSFIRE and LRIS & \citet{coil2004, masters2019, kriek2015}\\
 Keck II/DEIMOS & \citet{masters2019, cooper2012b, newman2013}\\
 HST/WFC3-IR grism & \citet{momcheva2016}
\enddata
\tablecomments{VLT: Very Large Telescope, VIMOS: Visible Multi-Object Spectrograph, FORS: the visual and near UV FOcal Reducer and low dispersion Spectrograph, IMACS: Inamori Magellan Areal Camera and Spectrograph, FMOS: Fiber multi-Object Spectrograph, LRIS: Low Resolution Imaging Spectrometer, MOIRCS: Multi-Object Infrared Camera and Spectrograph, MUSE: Multi Unit Spectroscopic Explorer, GMOS: Gemini Multi-Object Spectrographs, MOSFIRE: Multi-Object Spectrometer For Infra-Red Exploration, AAT: Anglo-Australian Telescope}

\label{table:spec_obs}
\end{deluxetable*}

\section{Sample Selection} \label{sec:obs_sfr_sampleselection}

In this section we describe the selection criteria we use to identify (i) the spectroscopic galaxy pair sample, (ii) the visually identified-interacting galaxy and merger samples, and (iii) the corresponding mass-, redshift-, and environment-matched isolated (control) galaxy samples. Further details can be found in \citet{shah2020}.

\subsection{Pair Selection}
We use both photometric and spectroscopic catalogs to obtain information about the positions, stellar masses, and best available spectroscopic redshifts of galaxies in the CANDELS and COSMOS fields. We use the following criteria to select our sample of massive spectroscopic galaxy pairs going through major galaxy interactions:

\begin{enumerate}

\item \textit{Redshift limit}: We require that both galaxies in a pair have reliable (quality flag greater than one) spectroscopic redshifts ($z_{spec}$) and that they span $0.5<z_{spec}<3$, enabling the inclusion of high-redshift interactions.

\item \textit{Mass limit}: The stellar mass of both of the galaxies in the pair has to be greater than 10$^{10}$\thinspace M$_{\odot}$ as this study focuses on massive galaxies within the completeness limits of the surveys.

\item \textit{Stellar mass ratio}:  The stellar mass ratio of the primary (more massive) to the secondary galaxy has to be less than four as this study focuses on major galaxy interactions.

\item \textit{Projected separation}: We require the projected separation between the two companion galaxies to be less than 150\thinspace kpc. Ideally, the three dimensional separation between the paired galaxies should be used to identify galaxy pairs. However, with observations we only have information about the projected separation between galaxies. We estimate the physical projected separation between two galaxies using their angular separation and average spectroscopic redshift. To constrain the line of sight separation, we use the relative radial velocities obtained using the spectroscopic redshifts of the galaxies. 

\item \textit{Relative line of sight velocity ($\Delta V$)}: Companions are required to have their relative line of sight velocity (obtained using their spectroscopic redshifts) within 5000\thinspace km s$^{-1}$. We also apply a stricter velocity cut of $\Delta V < 1000$\thinspace km s$^{-1}$ to study how the results vary with the relative line of sight velocity. We emphasize here that the likelihood of a pair being a true interaction increases as the relative velocity decreases. We include a wide range of $\Delta$V values in order to investigate the effect of different cuts. See \cite{shah2020} for a more detailed description and the $\Delta$V distribution of the full sample.

\end{enumerate}

We select the closest companion satisfying the criteria described above as the secondary galaxy corresponding to a given primary galaxy. Our full spectroscopic galaxy pair sample contains a total of 2351 pairs, while our sample with the stricter relative line of sight velocity cut of $\Delta V < 1000$\thinspace km s$^{-1}$, contains 1327 spectroscopic galaxy pairs.

\subsection{Visually Identified Interactions and Mergers}\label{sec:obs_sfr_visual_int_mer}

To study different stages of interactions and mergers, we also select sub-samples of visually identified interacting galaxies and mergers from the CANDELS fields using the classification scheme and catalog from \citet{kartaltepemor15}, with the constraints described in detail in \citet{shah2020}. Each galaxy was classified by at least three independent researchers.

We use three types of visually identified sub-samples: Merger, Blended Interaction, and Non-blended Interaction. A merger is a single coalesced system and an interaction is a system with at least two visually distinguishable galaxies. An interaction is considered blended if the photometric measurements correspond to both galaxies combined light. Galaxies in each of these samples are required to show signs of morphological disturbance such as tidal tails, loops, asymmetries, and off-center or highly irregular outer isophotes. Additionally, mergers can also show double nuclei and interactions can also show tidal bridges. In brief, the constraints for the different visually identified samples in the CANDELS fields are:

\begin{enumerate}
    \item H band magnitude of the galaxy has to be less than 24.5. This is a constraint for the \citet{kartaltepemor15} visual classification. 
    \item $\geq2/3$  of all classifiers agree on the classification of the system as a Merger, Blended interaction, or Non-blended interaction. 
    \item  The photometric redshift of each galaxy has to be between 0.5 and 3.0.
    
    \item The stellar mass of galaxies classified as a merger has to be greater than $1.25\times10^{10}$\thinspace M$_{\odot}$ (i.e., mass of the merger if two galaxies both with M$_{*}>10^{10}$\thinspace M$_{\odot}$ and a minimum stellar mass ratio of 0.25 merge together). Similarly,  the stellar mass of the blended interacting galaxy system and non-blended interacting galaxies has to be greater than $1.25\times10^{10}$\thinspace M$_{\odot}$ and 10$^{10}$\thinspace M$_{\odot}$,  respectively.
    
\end{enumerate}

In total, we compiled samples of 66 high-confidence mergers, 100 blended interactions, and 122 non-blended interactions in the CANDELS fields.

\subsection{Control Samples}\label{sec:controls_def}

The goal of this study is to estimate the effect of galaxy interactions on the SFRs of galaxies. However, the SFR of a galaxy can also vary with other galaxy properties, such as stellar mass, redshift, and environment. Hence, we generate a control sample of isolated galaxies with similar stellar mass, redshift, and environment as the paired galaxies and visually identified samples, and then compare the SFR of the pairs and control samples to estimate the effect of galaxy interaction on the SFRs of pairs.

The spectroscopic completeness varies significantly for each of different fields used in our analysis.
The availability of spectroscopic observations for a given galaxy can depend on properties such as its stellar mass, photometric redshift, SFR, AGN presence, etc. Furthermore, the spectroscopic redshift completeness also varies from field-to-field. Therefore, we choose pairs and controls from the same fields in order to avoid any bias based on the variation in spectroscopic redshift completeness in different fields. Like our paired galaxies, our control galaxies also must have a spectroscopic redshift with a quality flag greater than one. This requirement is to ensure that whatever biases are inherent in the spectroscopic redshift selection are present in both the pair and control samples. However, as we do not require a spectroscopic redshift for the visually identified interaction and merger samples, we also do not require spectroscopic redshifts for their controls.

We define the environmental overdensity as the ratio of the density at the position (RA, Dec, and redshift) of the galaxy to that of the median density in that redshift bin. For galaxies in the CANDELS fields, the overdensity measurement is based on the Monte Carlo Voronoi Tessellation method \citep{lemaux2017,tomczak2017}.  For the COSMOS field, the overdensity was derived based on redshift-dependent `weighted' adaptive kernel density maps generated by \citet{darvish2015}. \citet{darvish2015} also show that in spite of these two methods being slightly different from each other, their results are consistent. We use the methods mentioned above to estimate the density of the paired galaxies and control candidate galaxies and then consistently calculate the overdensity.

\begin{figure*}
    \centering
    \hspace{-0.2in}
    \includegraphics[trim = 2cm 5.9cm 1.3cm 4.0cm,clip=true, scale=0.7]{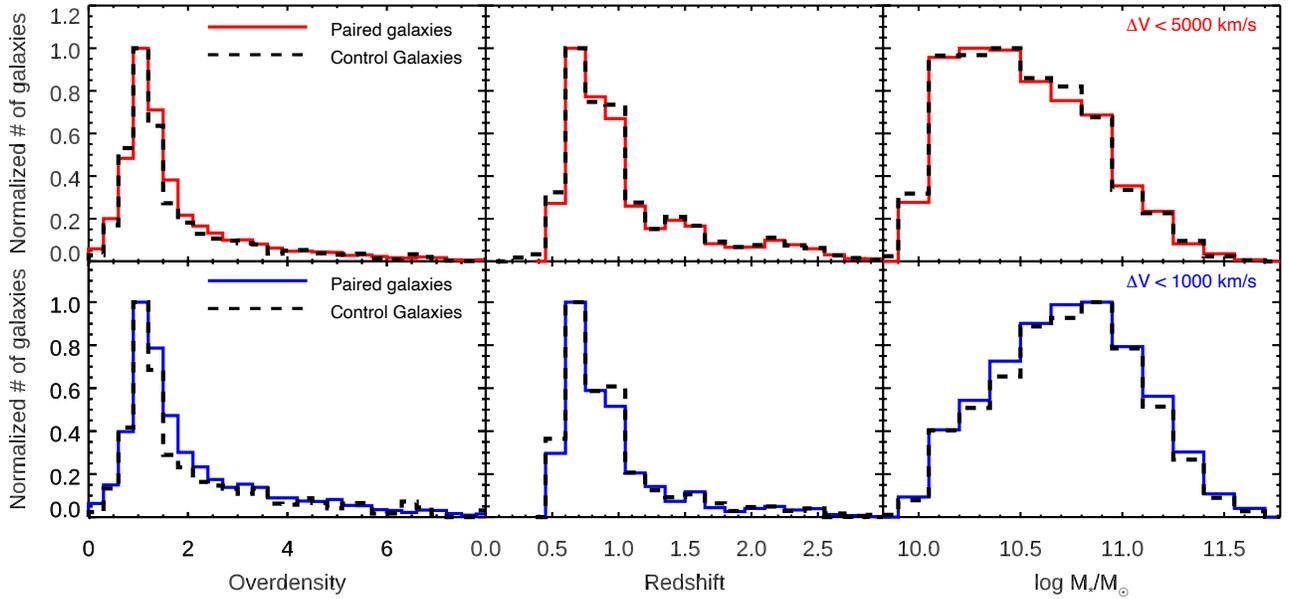}
    \caption{Normalized (at peak) distributions of environmental overdensity (left panel), spectroscopic redshift (middle panel),
and stellar mass (right panel) of spectroscopic galaxy pairs (solid lines) (with projected
separation $< 150$\,kpc, mass ratio $< 4$, and spectroscopic redshift between 0.5 and 3 for the $\Delta V < 5000$\thinspace km s$^{-1}$, top, and $\Delta V < 1000$\thinspace km s$^{-1}$, bottom, samples, respectively) and their corresponding mass-, redshift-, and environment-matched control galaxies (dashed black lines). }
    \label{fig:sfr_enh_distr_panel}
\end{figure*}
We generate a control candidate sample of isolated galaxies that have no major or minor companion (within a mass ratio of 10) within a $\Delta z$ corresponding to a relative velocity of less than 5000\thinspace km s$^{-1}$, out to a projected separation of 150\thinspace kpc. We also update the control candidate sample by removing visually identified interacting or merging galaxies (Section \ref{sec:obs_sfr_visual_int_mer}) from this control candidate sample.

We construct the final control sample by using this updated control candidate sample. As we plan to control for the effects of redshift, stellar mass, and environment on the SFRs of galaxies, the distribution of these properties for the control sample should be similar to the distribution of these properties for the paired galaxy sample. Therefore, we find the control galaxies for a given paired galaxy by minimizing the difference between these properties of the paired galaxy and control candidate galaxies. Hence, for each paired galaxy, we select three control candidate galaxies with the smallest  $ (\Delta\log{M_*})^2+(\Delta z)^2+(1/40)(\Delta overdensity)^2$ from the updated control candidate sample. Here, 1/40 is a weighting factor for the environment overdensity. As the stellar mass, redshift, and overdensity spans quite different ranges of values and have a different distribution, to best match in all three dimensions and avoid overdensity-matching dominating, we used this weighting factor (1/40). For the final control sample, the controls match within a stellar mass of 0.15\thinspace dex, spectroscopic redshift within 0.15, and overdensity within a factor of one for more than 90\% of the paired galaxies.

We show normalized distributions of the environmental overdensity, redshift, and stellar mass distribution of the final galaxy pair sample and corresponding control sample in Figure~\ref{fig:sfr_enh_distr_panel}. The plots show that the environmental overdensity, redshift, and stellar mass distributions for the pair sample and its control sample are very similar to each other, i.e., the controls are well-matched to the pair sample, which is crucial for our analysis \citep[see][]{shah2020}. Our samples span a wide range of environment, redshift, and stellar mass. The distributions in Figure~\ref{fig:sfr_enh_distr_panel} show that the number of galaxy pairs in our sample decreases rapidly at high overdensities and high redshift.

\section{Star formation Analysis}\label{sec:sf-analysis}

To estimate the star formation enhancement in merging and interacting galaxies, we first estimate the SFR of galaxies in the spectroscopic paired galaxy sample, visually identified interaction and merger galaxy samples, and the corresponding control galaxy samples as described in this section. We then compute the star formation enhancement and investigate how it varies with the galaxy pairs' projected separation. 

\subsection{Measurement of SFR}\label{subsec:sf-est}

We use the SED fitting tool Multi-wavelength Analysis of Galaxy Physical Properties \citep[{\sc MAGPHYS};][]{dacunha2008} for fitting the model SEDs to the photometric data points (FUV-FIR band flux and flux-error values) to measure the SFR and stellar mass of galaxies. We choose MAGPHYS for the SED fitting process as it self-consistently fits observations from UV to FIR based on an energy balance argument. It considers a combination of hot and cold dust grains as well as PAHs to estimate attenuation. MAGPHYS uses the \cite{bruzual2003} stellar population libraries. For estimating the SFR of the spectroscopic pairs and their corresponding control galaxies, we use their spectroscopic redshifts. For the visually identified interactions and merger galaxies and their corresponding control galaxies, we use their photometric redshifts.

The default version of MAGPHYS does not contain models with emission from an AGN component. However, it has been shown \citep[e.g.,][]{kirkpatrick2012} that a strong AGN can significantly impact the measured SFR of a galaxy, therefore, it is important to take such AGN activity into account. One way to identify galaxies that are strongly impacted by an AGN is to select galaxies that are poorly fit in the mid-infrared without the presence of an AGN component. Through substantial testing, it has been shown that the fitting residual at 8\thinspace $\mu$m can be indicative of strong AGN emission that dominates over star formation in the mid-infrared \citep{cooke2019}. Therefore, for galaxies for which the $8 \mu m$ residual percentile from the MAGPHYS fit ($100 \times \frac{flux_{obs}-flux_{MAGPHYS}}{flux_{obs}} \%$) is more than $40\%$, we use a modified version  of MAGPHYS called SED3FIT \citep{berta2013} that includes an AGN emission component to calculate their SFRs. We note that we elect not to fit the full sample using SED3FIT due to the computational time required and the fact that adding an additional component to the fit will often overestimate the importance of an AGN when there is not one present.  

\begin{figure*}
    \centering
    \includegraphics[trim = 2.3cm 2.7cm 3cm 2.8cm,clip=true, scale=0.38]{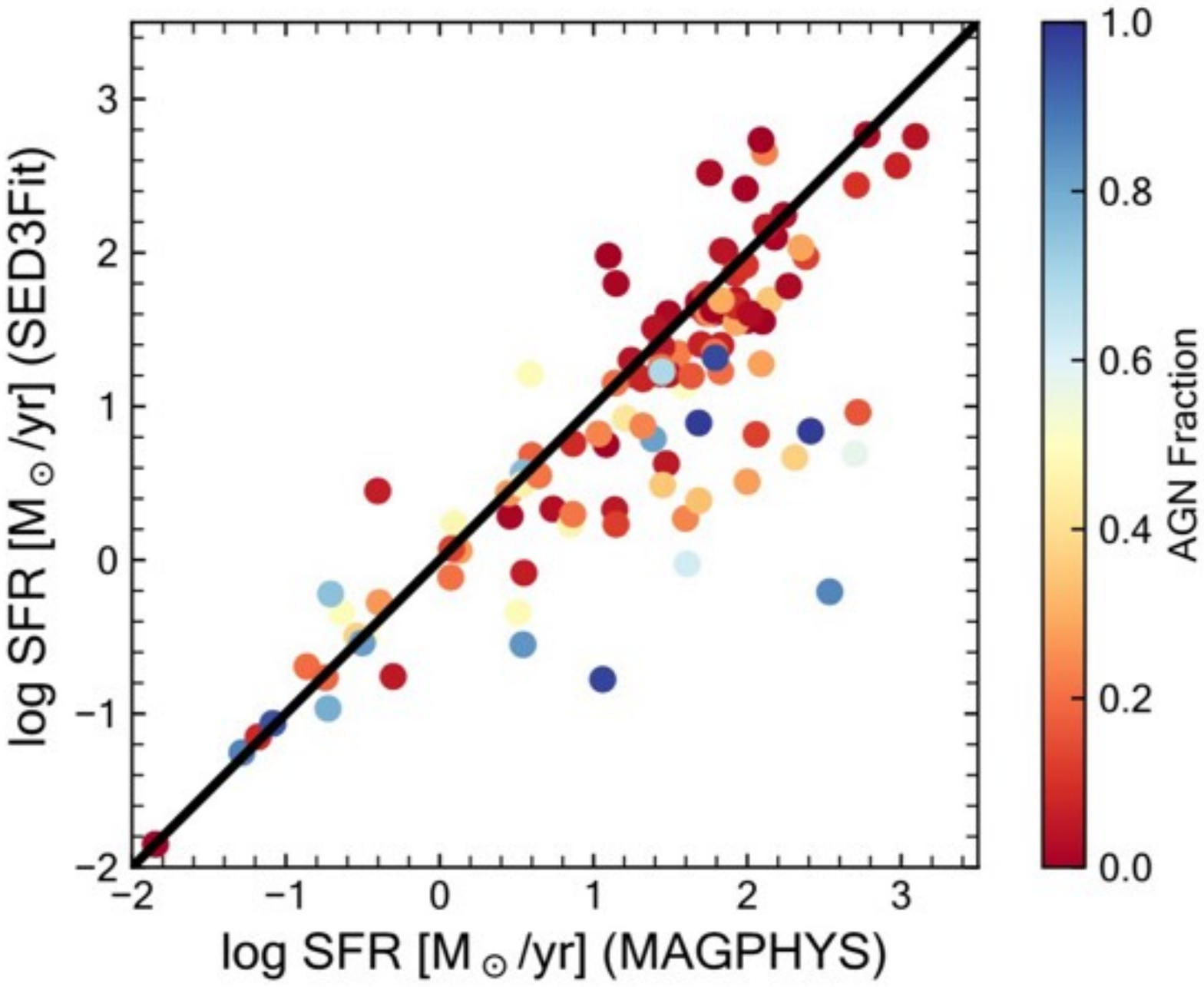}
        \includegraphics[trim = 2.3cm 2.76cm 3cm 2.5cm,clip=true, scale=0.37]{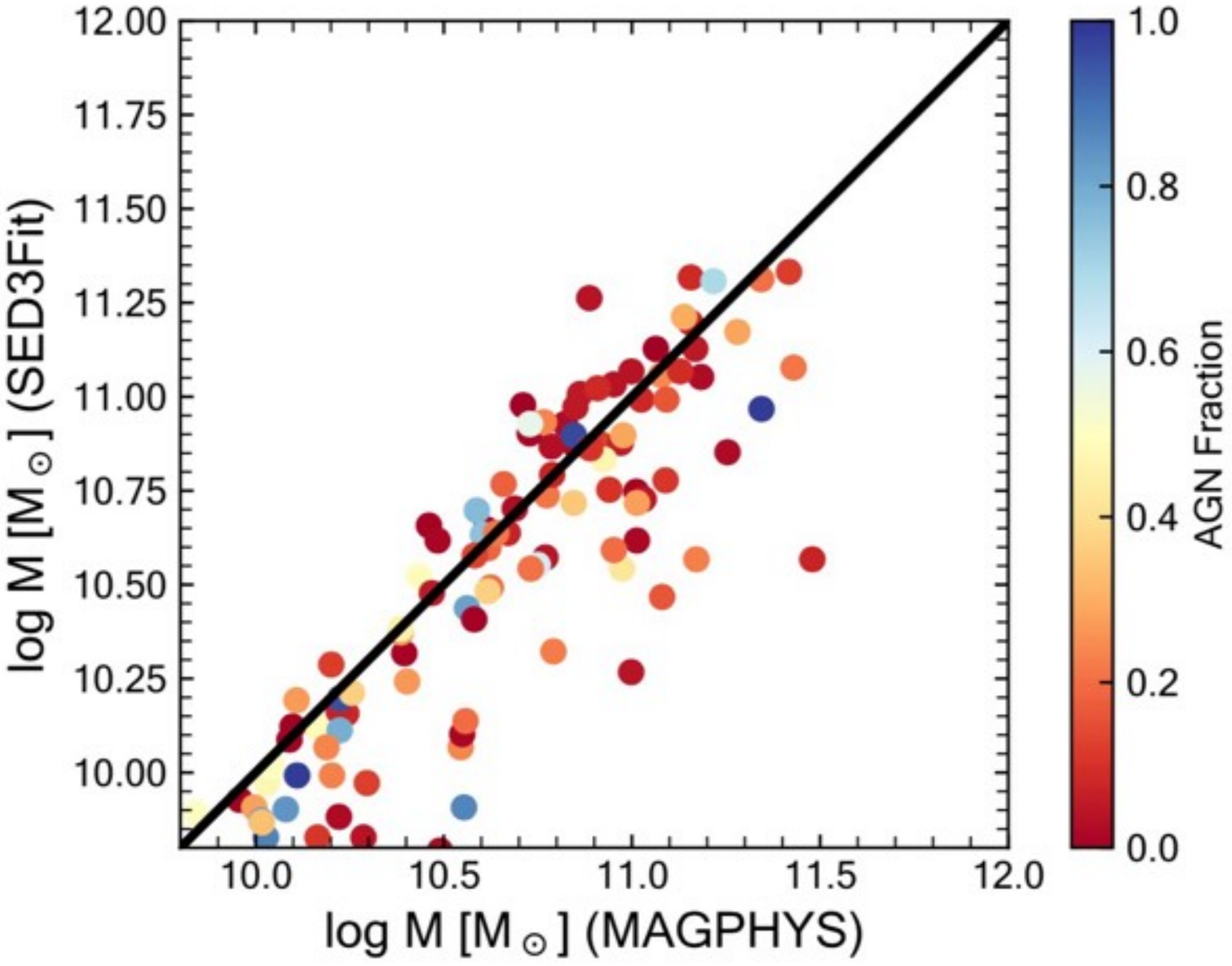}
    \caption{Comparison of the SFR (left) and stellar mass (right) measured using SED3FIT and MAGPHYS for galaxies in our sample with a large 8\,$\mu$m residual suggesting the presence of an AGN component in the mid-IR. The color bar shows the value of the AGN fraction estimated using SED3FIT for a given galaxy.}
    \label{fig:mstar_comp_bet_sed3_mag}
\end{figure*}

 \begin{figure*}
    \centering
    \includegraphics[scale=1.25]{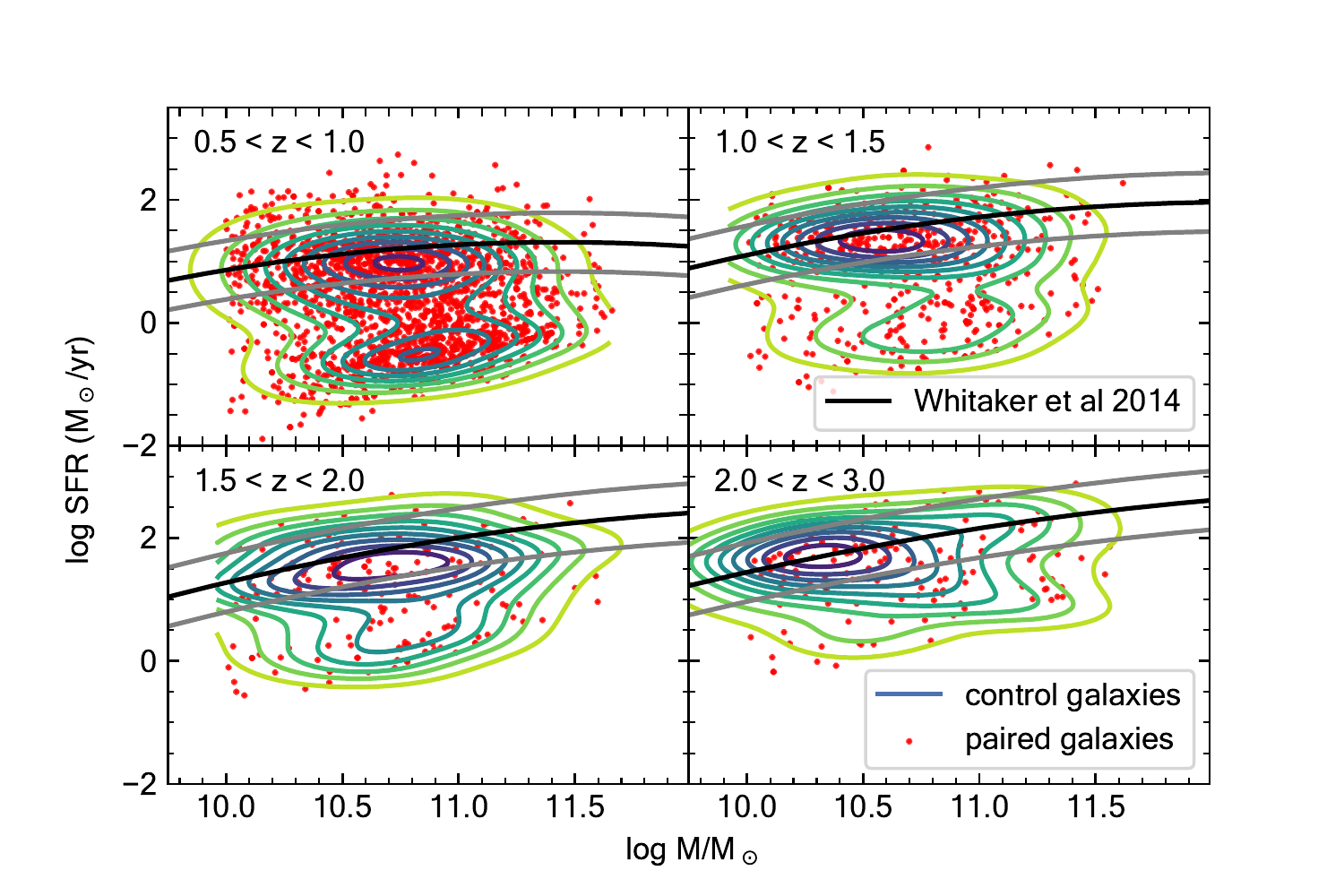}
    \caption{Distribution of paired galaxies (red scatter points) and controls (contours) on the SFR-M$_*$ plane for (top to bottom) the $0.5<z<3.0$, $0.5<z<1.0$, $1.0<z<1.5$, $1.5<z<2.0$, and $2.0<z<3.0$ bins. The black solid line in each sub-plot corresponds to the star formation main sequence \citep{whitaker2014} in the given redshift range. The gray lines correspond to the SFR value above and below the SFMS by a factor of three in each  redshift bin.}
    \label{fig:sfrms_z_con_contour}
\end{figure*}

\begin{figure*}
    \centering
    \includegraphics[scale=1.0]{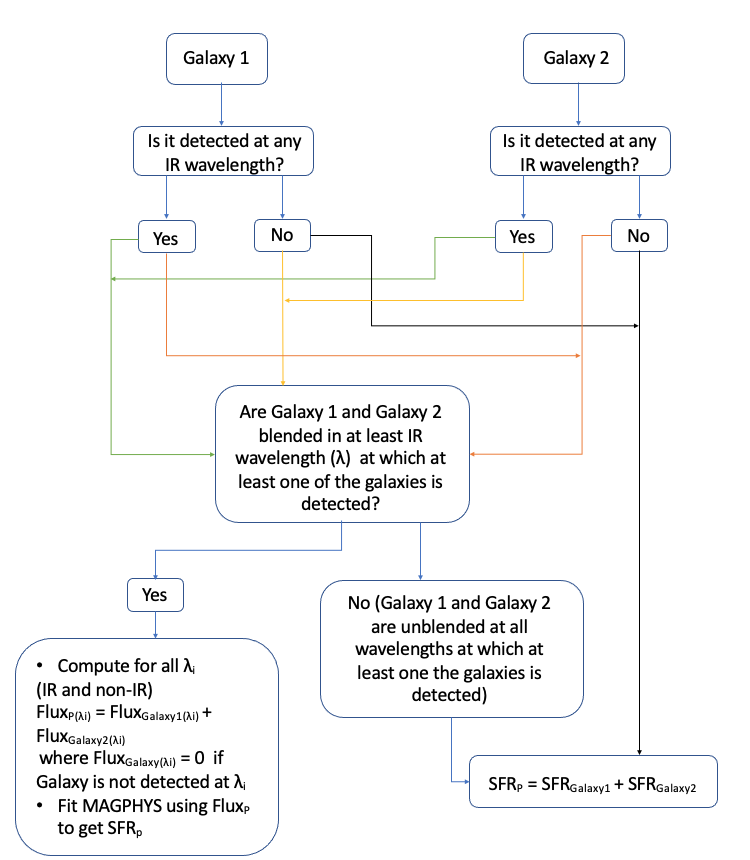}
    \caption{The decision tree for computing the combined SFR, SFR$_{P}$, of a galaxy pair, where galaxy1 and galaxy2 correspond to the primary and secondary galaxies in a given galaxy pair, respectively.}
    \label{fig:sfr_blend}
\end{figure*}

Figure~\ref{fig:mstar_comp_bet_sed3_mag} shows the comparison of the SFRs measured using SED3FIT and MAGPHYS. For most galaxies, the SFRs measured using the two SED fitting routines are similar to each other. However, for many galaxies the SFR measured with SED3FIT is lower than the value from MAGPHYS. This is mainly due to the fact that MAGPHYS can overestimate the total SFR when a strong AGN is present. Similarly, the right panel of Figure~\ref{fig:mstar_comp_bet_sed3_mag} shows a comparison between the stellar mass measured using SED3FIT and MAGPHYS for the same subset of galaxies with a large 8\,$\mu$m residual. Again, for most galaxies, the stellar masses measured using SED3FIT are similar or lower than that from MAGPHYS. For these objects, we adopt the SED3FIT masses and SFRs in order to avoid overestimates due to the presence of a strong AGN. We note that these strong AGN are rare, and therefore, have minimal impact on our final results. For consistency, we remove any pairs from our sample that no longer satisfy the selection criteria as described in \S~\ref{sec:obs_sfr_sampleselection} with these new stellar masses. This results in the removal of 17 pairs for the $\Delta V < 1000$\thinspace km s$^{-1}$ sample and 22 pairs for the $\Delta V < 5000$\thinspace km s$^{-1}$ sample. These removed pairs are reflected in the final pair sample presented in \S~\ref{sec:obs_sfr_sampleselection} and the figures throughout the paper. 

 Using the results of our SED fits, we show the distribution of paired and control galaxies on the SFR-M$_*$ plane in different redshift bins in Figure \ref{fig:sfrms_z_con_contour}. We also show the corresponding star-forming main sequence \citep{whitaker2014} at each redshift. The two peaks in the contours in each plot correspond to the main sequence and the quiescent galaxy populations. Overall the control contours seem to trace the paired galaxy population well. The number of objects decreases rapidly with increasing redshift due to spectroscopic redshift incompleteness (especially at high redshifts). Note that the galaxy pair population spans a wide range of properties, including starbursts and quiescent galaxies, above and below the main sequence, respectively. The paired galaxies are required to have a minimum stellar mass of $10^{10}$ M$_\odot$, however the controls can have a slightly lower stellar mass than $10^{10}$ M$_\odot$ and so the contours extend slightly to the left of the paired galaxies in some of the sub-plots.

\subsection{Blending of paired galaxies at IR wavelengths}
The angular resolution of the images taken at 24\thinspace $\mu$m and longer wavelengths is poor relative to the optical/NIR filters and therefore, some galaxies in close pair systems are blended at MIR-FIR wavelengths. This means that for such pairs that are blended only at longer wavelength, the corresponding flux values will represent the pair and not the individual galaxies, which will affect their SFR measurement. Therefore, the flux values of the whole system, i.e., both galaxies combined, have to be used across all wavelengths when fitting an SED of the blended pair system in order to obtain an accurate measurement.

We compute the SFR of the whole galaxy pair system (both galaxies combined, SFR$_{P}$) for all galaxy pairs for a fair comparison with systems blended in the IR. Figure \ref{fig:sfr_blend} illustrates the decision tree for measuring SFR$_{P}$  for a given galaxy pair. We start with both galaxies (primary: galaxy1 and secondary: galaxy2) in the pair and determine if either of them are detected in any IR band. If neither of the two galaxies in a given galaxy pair are detected in any IR bands (black arrows), then the combined SFR$_{P}$ is the summation of the SFR of the individual galaxies. If at least one of the galaxies in the pair is detected in at least one of the IR bands (say $\lambda$), then we check if the galaxies are blended at that wavelength $\lambda$. We repeat this process of checking blending for all wavelengths ($\lambda$1, $\lambda$2,...,$\lambda$n) in which at least one of the two galaxies are detected. If both galaxies are not blended at all wavelengths, we compute the SFR$_{P}$ by adding the SFR of the two individual galaxies. If the two galaxies are blended in at least one of the wavelengths ($\lambda$ in $\lambda$1, $\lambda$2,...,$\lambda$n) then we combine the fluxes of the two galaxies at each wavelength to get the combined flux of the system in each band. We then fit the SED of this combined system using MAGPHYS to obtain SFR$_{P}$. If the $8\, \mu m$ percentile residual of the MAGPHYS fit for the combined system is more than 40 $\%$, then we refit the SED using SED3FIT to obtain the SFR$_{P}$. Only 115 galaxy pairs, i.e, $\sim 8.5\%$ of our total galaxy pair sample, are blended in the IR.

To facilitate a fair comparison between the pair system and their control counterparts, we compute the mean SFR of all controls of one of the two galaxies in a given pair ($\overline{SFR_{C1}}$) and repeat the process for the second galaxy ($\overline{SFR_{C2}}$). We then add these mean SFRs to obtain the final control SFR for a given galaxy pair ($SFR_{C} = \overline{SFR_{C1}} + \overline{SFR_{C2}}$). We propagate the errors in SFRs throughout this process to get the upper and lower error estimates for the SFR of combined control SFR ($SFR_{C}$) for a given galaxy pair. We repeat the process for all galaxy pairs. 

\section{SFR Enhancement}\label{sec:sfr_enh}

We define the SFR enhancement as the ratio of the weighted mean of the SFR of all galaxy pairs and the weighted mean SFR of all the corresponding combined controls, i.e., 

\begin{equation}\label{eq:enh}
    SFR\: Enhancement = \frac{\overline{SFR_{P}}}{\overline{SFR_{C}}}.
\end{equation}

\begin{figure}
    \centering
    \includegraphics[trim = 2.5cm 1.4cm 2.4cm 2.3cm,clip=true, scale=0.385]{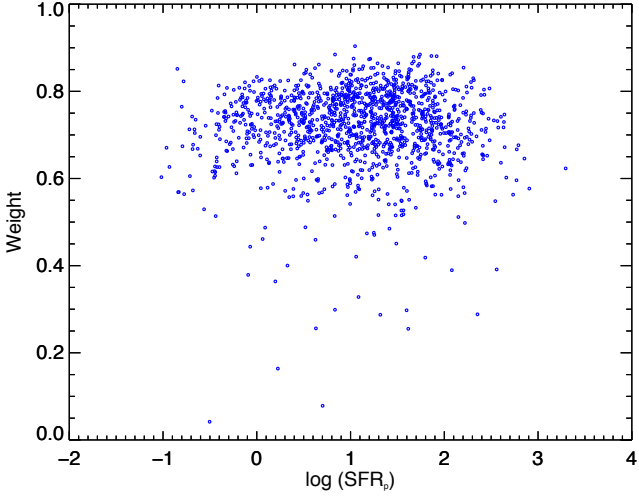}
    \caption{Distribution of weights (w$_{i}$) with respect to the SFR of galaxy pairs ($SFR_{P}$). The distribution shows that there is no significant trend between the weights and $SFR_{P}$.}
    \label{fig:weight_vs_sfr_scatter}
\end{figure}

For some galaxies, there are large errors on the SFR measured from our SED fits, which could be due to various reasons such as a poor fit, limitation of number of models in a given parameter space, large errors on the observed fluxes, etc. We consider this fact while estimating the average SFR of the pair sample ($SFR_{P}$) and the average SFR of the combined control sample (SFR$_{C}$). To decrease the dependence of the results on systems with large errors in the SFR measurement, we multiply the SFR$_p$ and SFR$_C$ values by a weighting function such that a lower weight value is assigned for a larger SFR percentile error and a higher weight value is assigned for a lower SFR percentile error. For a given galaxy pair, we use a weighting factor (w) based on the combined error (x) for SFR$_{P}$ and $SFR_{C}$ using the formulae
\begin{equation} \label{eq:xl}
    x_{low} = \sqrt{\left(\frac{(\Delta SFR_{P})_{low}}{SFR_{P}}\right)^2 + \left(\frac{(\Delta SFR_{C})_{low}}{SFR_{C}}\right)^2},
\end{equation}

\begin{equation} \label{eq:xu}
    x_{up} = \sqrt{\left(\frac{(\Delta SFR_{P})_{up}}{SFR_{P}}\right)^2 + \left(\frac{(\Delta SFR_{C})_{up}}{SFR_{C}}\right)^2},
\end{equation}

\begin{equation} \label{eq:x}
    x = \sqrt{\left(\frac{x_{low}^2 + x_{up}^2}{2}\right)},
\end{equation}
and
\begin{equation}\label{eq:weight}
   w = \frac{1}{1+x}.
\end{equation}

\noindent Here, $(\Delta SFR_{P})_{low}$ and $(\Delta SFR_{P})_{up}$ are the lower and upper errors on the SFR of the given pair ($SFR_{P}$). Similary, $(\Delta SFR_{C})_{low}$ and $(\Delta SFR_{C})_{up}$ are the lower and upper errors on the SFR of the combined controls for a given galaxy pair ($SFR_{C}$). We also checked that our weights are not biased towards galaxy properties such as their M$_*$ and SFR. Our enhancement values do not vary significantly if we use different weighting functions. 
For a given galaxy pair (for example, i), we compute the value of the combined error (x$_{i}$) using Eq.\ \ref{eq:xl}, \ref{eq:xu}, and \ref{eq:x}. We then use the value of x$_{i}$ to calculate the value of the weight (w$_{i}$) for the given galaxy pair using Eq.\ \ref{eq:weight}. The weighting function values (w$_{i}$) decrease with increasing average percentage error (x$_{i}$) on the SFR. We show the distribution of these weights (w$_{i}$) with respect to the SFR of galaxy pairs ($SFR_{P}$) in Figure~\ref{fig:weight_vs_sfr_scatter}. The figure shows that there is no significant trend between the SFR and the weights. Hence, by using the weights, our analysis is not biased towards higher or lower SFRs.

\begin{figure}
    \centering
    \includegraphics[trim = 4.0cm 3.9cm 4.0cm 5.4cm,clip=true, scale=0.44]{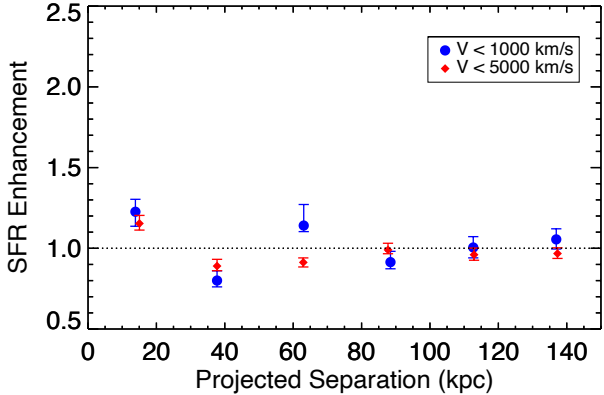}
    \caption{SFR enhancement as a function of projected separation of pairs. SFR enhancement is defined as the ratio of the weighted average of the SFR of the pair sample to that of the control sample (Equation \ref{eq:enh}). The filled blue circles and filled red diamonds correspond to the SFR enhancement for the $\Delta V<1000$\thinspace km s$^{-1}$ and $\Delta V<5000$\thinspace km s$^{-1}$  pair samples, respectively. The dotted line corresponds to a value of one, which represents no interaction-induced SFR enhancement. We see the highest level of SFR enhancement at a factor of 1.23$^{+0.08}_{-0.09}$ ($\sim2.6\sigma$) and  1.15$^{+0.05}_{-0.04}$ ($\sim3.0\sigma$) in the closest separation bin with projected pair separation $<$25\thinspace kpc for galaxy pair sample with $\Delta V < 1000$\thinspace km\thinspace s$^{-1}$ and $\Delta V < 5000$\thinspace km\thinspace s$^{-1}$, respectively. We note that there is a significant amount of scatter around the SFR enhancement value of one. }
    \label{fig:sfr_enh_pairs}
\end{figure}

Finally, we compute the weighted average of the SFR for the galaxy pair sample and its corresponding control samples using

\begin{equation}\label{eq:sfrp}
  \overline{SFR_{P}}  = \frac{\sum_i^{n_{p}} w_i  (SFR_{P})_i}{\sum w_i}
\end{equation} and

\begin{equation}\label{eq:sfrc}
  \overline{SFR_{C}}  = \frac{\sum_i^{n_{p}} w_i (SFR_{C})_i}{\sum w_i} .
\end{equation}

\begin{figure*}
    \centering
    \includegraphics[trim = 2cm 3.5cm 3cm 10cm,clip=true, scale=0.75]{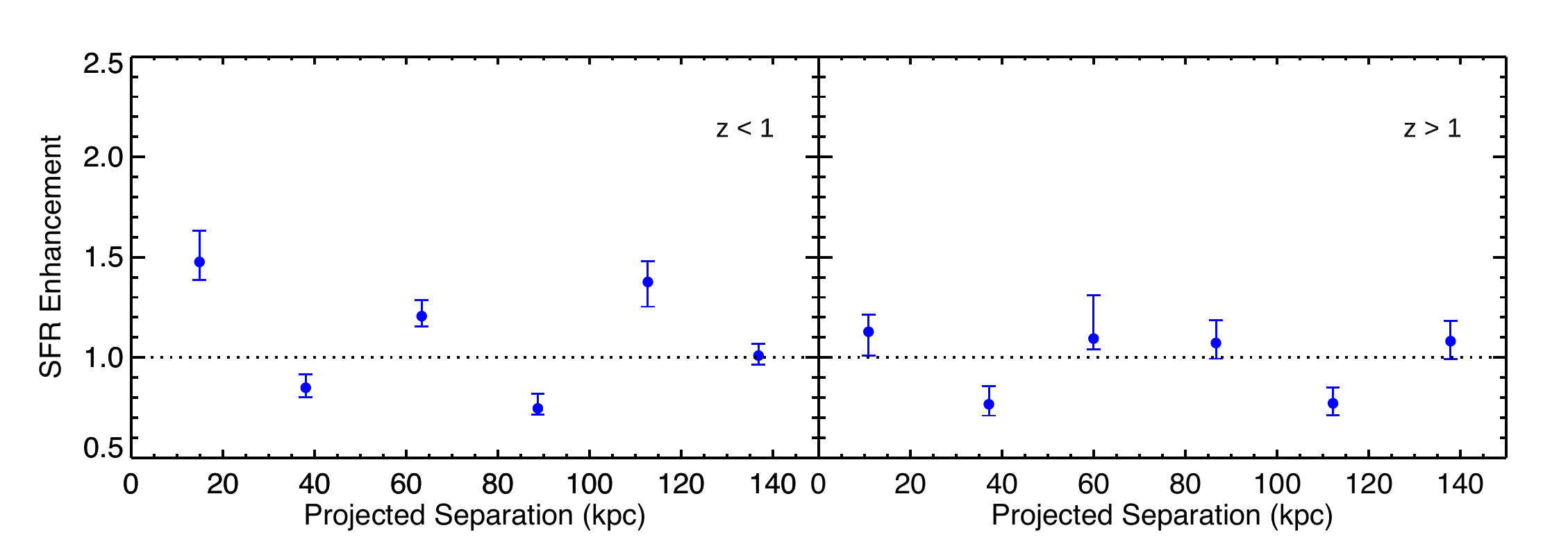}
    \caption{SFR enhancement as a function of projected separation for the galaxy pairs with $\Delta V < 1000$\thinspace km\thinspace s$^{-1}$ at $z<1$ (left) and $z>1$ (right). The dotted line corresponds to a value of one, which represents no interaction-induced SFR enhancement.}
    \label{fig:sfr_enh_zdiv}
\end{figure*}

\begin{figure*}
    \centering
    \includegraphics[trim = 2cm 3.5cm 3cm 10cm,clip=true, scale=0.75]{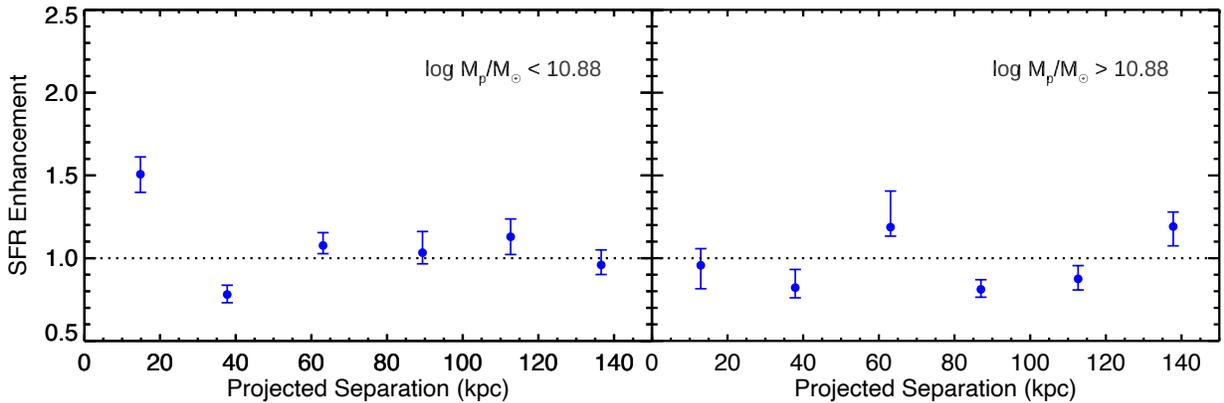}
    \caption{SFR enhancement as a function of projected separation for the galaxy pairs with $\Delta V < 1000$\thinspace km\thinspace s$^{-1}$ and with the stellar mass of the primary (more massive) galaxy $M_{p}$ less than 10.88\thinspace M$_{\odot}$ (left) and more than 10.88\thinspace M$_{\odot}$ (right) for the pair sample with $\Delta V < 1000$\thinspace km\thinspace s$^{-1}$. The dotted line corresponds to a value of one, which represents no interaction-induced SFR enhancement.}
    \label{fig:sfr_enh_mdiv}
\end{figure*}

\begin{figure*}
    \centering
    \includegraphics[trim = 2cm 2.0cm 3cm 10cm,clip=true, scale=0.75]{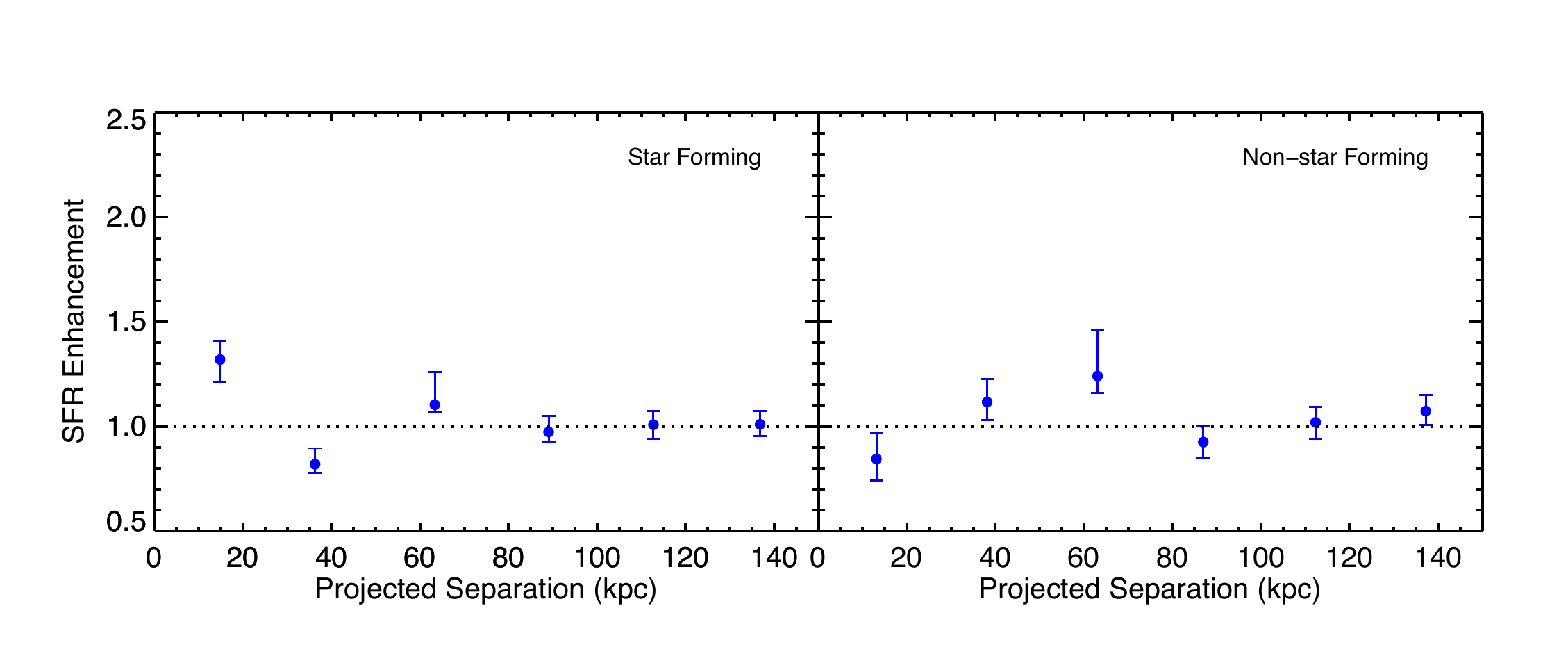}
    \caption{SFR enhancement as a function of projected separation for the pair sample with $\Delta V < 1000$\thinspace km\thinspace s$^{-1}$.     The left and right hand side plots correspond to the results for star-forming blended pairs (and non-blended paired galaxies) and non-star forming blended pairs (and non-blended paired galaxies). The dotted line corresponds to a value of one, which represents no interaction-induced SFR enhancement.}
    \label{fig:sfr_enh_sfdiv}
\end{figure*}

We use the weighted average $\overline{SFR_{P}}$ and $\overline{SFR_{C}}$ values to estimate the SFR enhancement using Eq.~\ref{eq:enh}. We divide the projected separation range ($0-150\thinspace$ kpc) into six different bins of width 25\thinspace kpc. We estimate the SFR enhancement in each bin. We show the SFR enhancement results for the spectroscopic galaxy pair samples with relative line of sight velocity $\Delta V < 1000$\thinspace km\thinspace s$^{-1}$ and $\Delta V < 5000$\thinspace km\thinspace s$^{-1}$ in Figure~\ref{fig:sfr_enh_pairs}. We see the highest level of SFR enhancement of a factor of 1.23$^{+0.08}_{-0.09}$ ($\sim2.6\sigma$) and  1.15$^{+0.05}_{-0.04}$ ($\sim3.0\sigma$) in the closest separation bin of projected pair separation $<$25\thinspace kpc for the galaxy pair sample with $\Delta V < 1000$\thinspace km\thinspace s$^{-1}$ and $\Delta V < 5000$\thinspace km\thinspace s$^{-1}$, respectively. We note that there is significant scatter around the SFR enhancement value of one (a value indicating no enhancement). The results also do not show a clear trend of increasing SFR enhancement with decreasing projected separation. While the error bars on the SFR enhancements obtained using the weighted averages as described above are smaller than the ones calculated without weighting, the overall value of the SFR enhancement calculated with and without weights are very similar. We also note that we do not see a large difference between the $\Delta V < 1000$\thinspace km\thinspace s$^{-1}$ and $\Delta V < 5000$\thinspace km\thinspace s$^{-1}$ samples. Since there are more galaxies with $\Delta V < 5000$\thinspace km\thinspace s$^{-1}$, the error bars and resulting amount of scatter among the points are smaller. However, since pairs with $\Delta V < 1000$\thinspace km\thinspace s$^{-1}$ are more likely to be physically associated, we adopt this subsample for the remaining plots and discussion.

We then divide the $\Delta V < 1000$\thinspace km\thinspace s$^{-1}$ galaxy pair sample ($z_{median}\sim1$) into two redshift bins ($z<1$ and $z>1$), with equal numbers of pairs in each bin, and calculate the SFR enhancement (Figure~\ref{fig:sfr_enh_zdiv}). We do not see a trend of increasing SFR enhancement with decreasing projected separations in either redshift bin. We see the highest enhancement of 1.48$^{+0.16}_{-0.09}$ in the lowest projected separation bin ($<25$\thinspace kpc) at $z<1$. We see a large amount of scatter around the SFR enhancement value of one for both redshift bins. This scatter is most likely due to the sample size and does not represent a real trend in the SFR enhancement levels with separation.

Similarly, we also divide our $\Delta V < 1000$\thinspace km\thinspace s$^{-1}$ galaxy pair sample into two stellar mass bins separated at the median stellar mass  ($10^{10.88}$\,M$_{\odot}$) of the primary galaxy (more massive of the two galaxies) in a galaxy pair and represent the results in Figure~\ref{fig:sfr_enh_mdiv}. In the smallest projected separation bin, we see a SFR enhancement of 1.51$^{+0.11}_{-0.10}$ ($\sim4.6\sigma$) in the lower mass sample, which is $\sim 1.63 \times$ higher (at $\sim5.4\sigma$ level) than the SFR enhancement of 0.96$^{+0.10}_{-0.14}$ in the same projected separation bin for the higher mass sample (log\thinspace$M_{prim}/M_{\odot}>10.88$). Again, with this subdivided sample, we see a significant amount of scatter around the SFR enhancement value of one. 

Furthermore, we divide the $\Delta V < 1000$\thinspace km\thinspace s$^{-1}$ galaxy pair sample into star-forming (SF) paired galaxies and non-star forming (non-SF/quiescent) paired galaxies identified based on whether SFR of the galaxy is higher or lower than the specific SFR of $sSFR=(1+z)^{2.5}\times10^{-11}$ \citep{speagle2014}, respectively. In the case of blended pairs, we apply this criterion to the sSFR of the whole blended system. We also select new sets of SF (non-SF) controls for this SF (non-SF) blended pairs and non-blended paired galaxies using the same method described previously. We show the SFR enhancement results in Figure~\ref{fig:sfr_enh_sfdiv}. In the closest separation bin, we see a SFR enhancement of 1.32$^{+0.09}_{-0.11}$ in the SF sample, which is $\sim 1.57\times$ larger than for the non-SF sample (0.84$^{+0.12}_{-0.10}$). Here also, the results show a significant amount of scatter around the SFR enhancement value of one. We also compare the SFR enhancements in our primary galaxy sample to the secondary galaxy sample of the non-blended galaxy pairs. We do not see a significant difference in SFR enhancement of these two samples (not shown here).

\begin{figure}
    \centering
    \includegraphics[trim = 2.3cm 3cm 3cm 4cm,clip=true, scale=0.38]{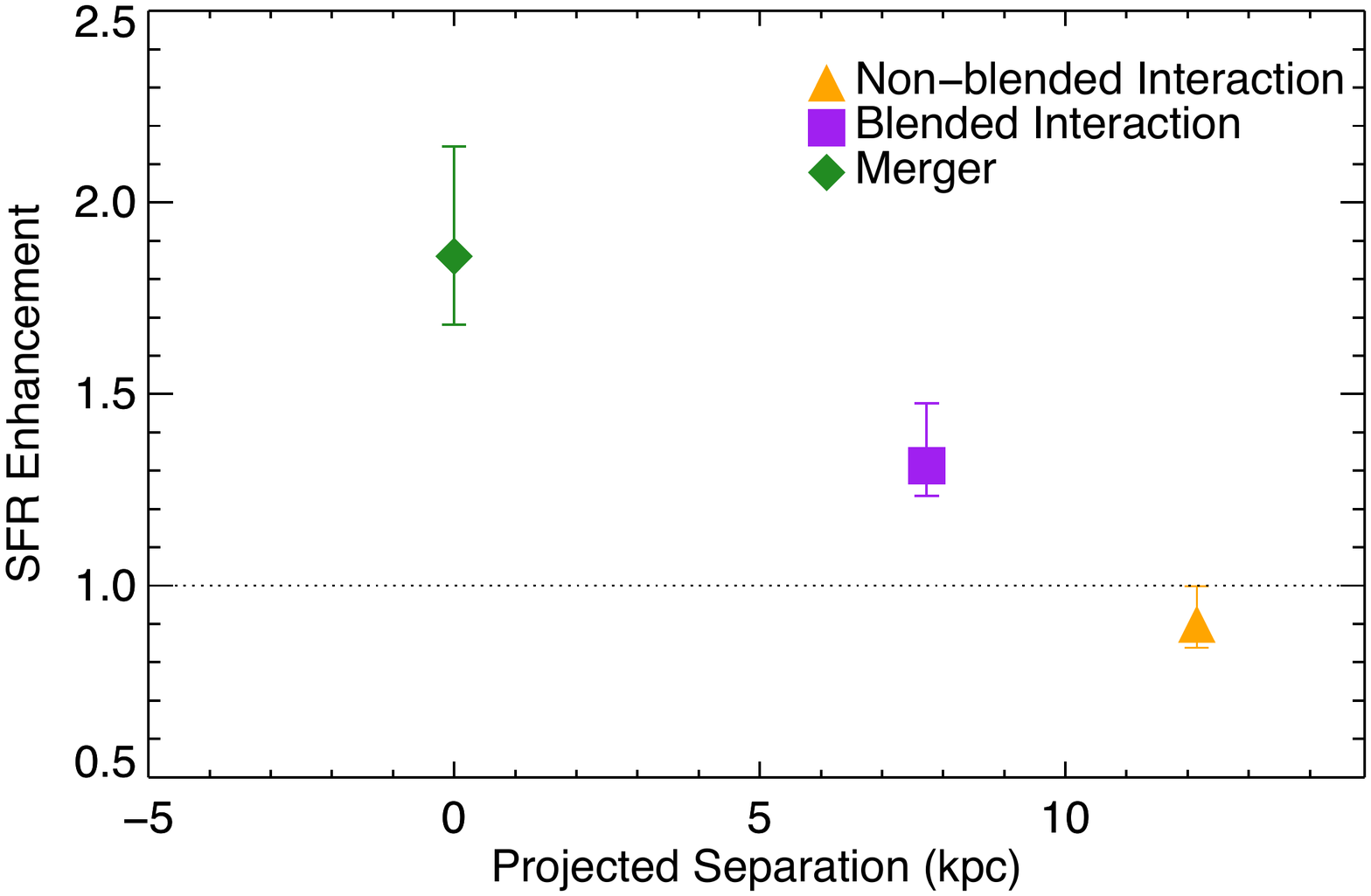}
    \caption{SFR enhancement for our visually identified mergers (filled green diamonds), blended interactions (filled purple squares), and non-blended interactions (filled orange triangles) as a function of their median projected separations. The merged/coalesced systems are plotted at a separation of zero, since they are no longer in a pair but are in a single coalesced galaxy. The dotted line corresponds to a value of one, which represents no interaction-induced SFR enhancement. Note that for this sample, we see a clear trend of increasing SFR enhancement with decreasing projected separation, with the largest level of enhancement seen in the coalesced systems.}
    \label{fig:sfr_enh_ls}
\end{figure}

We present the SFR enhancement for our visually identified interactions and merger samples in Figure~\ref{fig:sfr_enh_ls}. We see enhancements of 1.86$^{+0.29}_{-0.18}$, 1.31$^{+0.16}_{-0.08}$, and 0.90$^{+0.10}_{-0.06}$ for our visually identified merger sample, blended-interaction sample, and non-blended interaction sample, respectively. For this sample, we see a clear trend of increasing SFR enhancement for decreasing median projected separation, with the largest enhancement seen for coalesced merger systems.

\begin{figure*}
    \centering
    \includegraphics[trim = 2cm 2cm 3cm 10cm,clip=true, scale=0.75]{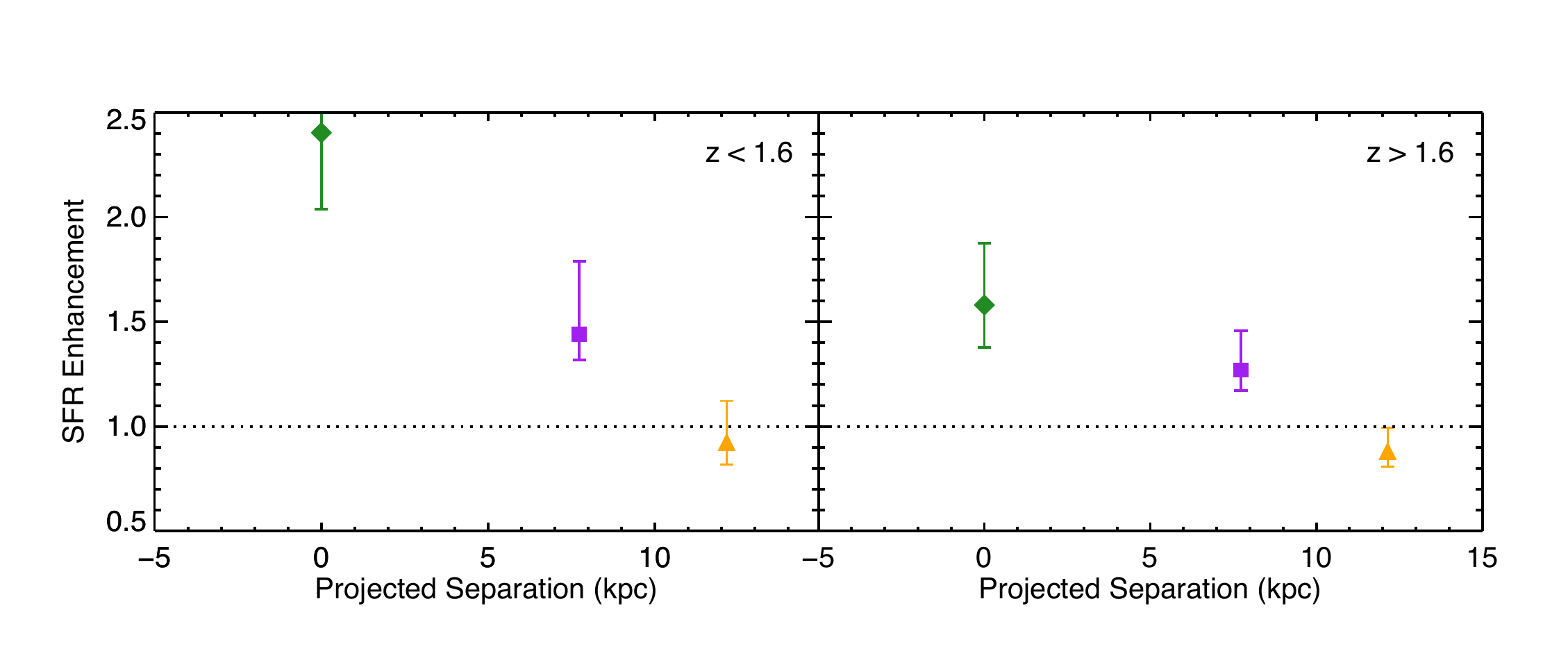}
    \caption{SFR enhancement for our visually identified mergers (filled green diamonds), blended interactions (filled purple squares), and non-blended interactions (filled orange triangles) at $z<1.6$ (left) and $z<1.6$ (right) compared to their star-forming control samples. The dotted line corresponds to a value of one, which represents no interaction-induced SFR enhancement. While we see evidence of increasing enhancement with decreasing projected separation in both redshift bins, we see a decreased level of enhancement for the $z>1.6$ bin.}
    \label{fig:sfr_enh_ls_zdiv}
\end{figure*}

Since we use photometric redshift for the visually identified sample, rather than spectroscopic redshifts as for the galaxy pair sample, the median redshift of these galaxies is higher than that of the pair sample. We divide our visually identified interaction and merger samples into two redshift bins separated at the median redshift $z_{median}\sim1.6$ of the samples. We show these results in Figure~\ref{fig:sfr_enh_ls_zdiv}. At low $z$ ($0.5<z<1.6$),  we find that the SFR enhancement levels are a factor of  2.40$^{+0.62}_{-0.37}$, 1.44$^{+0.35}_{-0.12}$, and 0.93$^{+0.20}_{-0.11}$ for our merger, blended, and non-blended interaction samples, respectively. Similarly, at high $z$ ($z>1.6$), we see enhancement values of 1.58$^{+0.29}_{-0.20}$, 1.27$^{+0.19}_{-0.10}$, and 0.88$^{+0.11}_{-0.08}$ for our merger, blended, and non-blended interaction samples, respectively. In both redshift bins, we find that there is a clear trend of increasing SFR enhancement with decreasing projected separation with the largest enhancement see for the coalesced merger systems. We also find that the level of enhancement at $z<1.6$ is a factor of $\sim 1.5$ higher than the enhancement at $z>1.6$, suggesting that there is an evolution in the interaction-induced SFR enhancement between these two bins.

\begin{figure*}
    \centering
    \includegraphics[trim = 2cm 2cm 3cm 10cm,clip=true, scale=0.75]{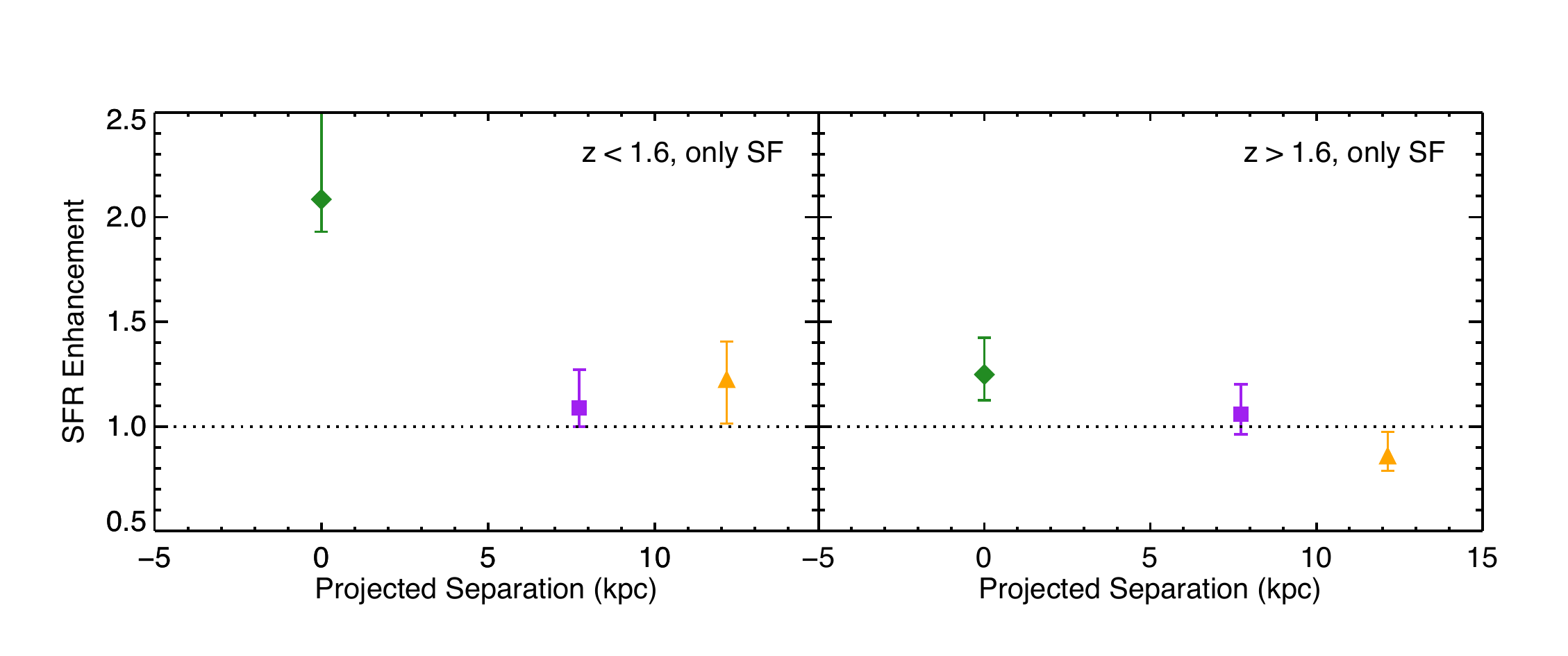}
    \caption{SFR enhancement for only star-forming visually identified mergers (filled green diamonds), blended interactions (filled purple squares), and non-blended interactions (filled orange triangles) at $z<1.6$ (left) and $z<1.6$ (right) compared to their star-forming control samples.}
    \label{fig:sfr_enh_ls_only_sf_zdiv}
\end{figure*}
Finally, we select the subset of galaxies in the visually identified sample that are star-forming using the same criteria as for the pair sample ($sSFR>(1+z)^{2.5}\times10^{-11}$; \citealt{speagle2014}) and selected their corresponding star-forming control samples. As shown in Figure~\ref{fig:sfr_enh_ls_only_sf_zdiv}, we calculate a SFR enhancement of 2.08$^{+0.42}_{-0.15}$, 1.09$^{+0.18}_{-0.09}$, and 1.23$^{+0.18}_{-0.21}$ at low $z$ for our merger, blended, and non-blended interaction samples, respectively. At high z, the corresponding SFR enhancement values are 1.23$^{+0.18}_{-0.12}$, 1.06$^{+0.15}_{-0.10}$, and 0.86$^{+0.11}_{-0.07}$.  We find that for the coalesced systems, the SFR enhancement if a factor of $\sim 1.7$ times higher at $z<1.6$ than at $z>1.6$.

\section{Discussion} \label{sec:obs_sfr_discussion}

We investigate the level of SFR enhancement in major spectroscopic galaxy pairs relative to their stellar mass-, redshift-, and environment-matched control sample of isolated galaxies at $0.5<z<3.0$. We find that there is a slight SFR enhancement of a factor of 1.23$^{+0.08}_{-0.09}$ ($\sim2.6\sigma$) and  1.15$^{+0.05}_{-0.04}$ ($\sim3.0\sigma$) in the lowest projected separation bin ($<25$\thinspace kpc) for our $\Delta V < 1000$\thinspace km\thinspace s$^{-1}$ and $\Delta V < 5000$\thinspace km\thinspace s$^{-1}$ spectroscopic galaxy pair samples, respectively. We find a stronger level of enhancement of a factor of 1.86$^{+0.29}_{-0.18}$ in the coalesced visually identified mergers. We compare these result with similar studies in the literature. 

\subsection{Comparison with Studies in the Local Universe}

\begin{figure*}
    \centering
    \includegraphics[trim = 2.2cm 3cm 3.4cm 4cm,clip=true, scale=0.5]{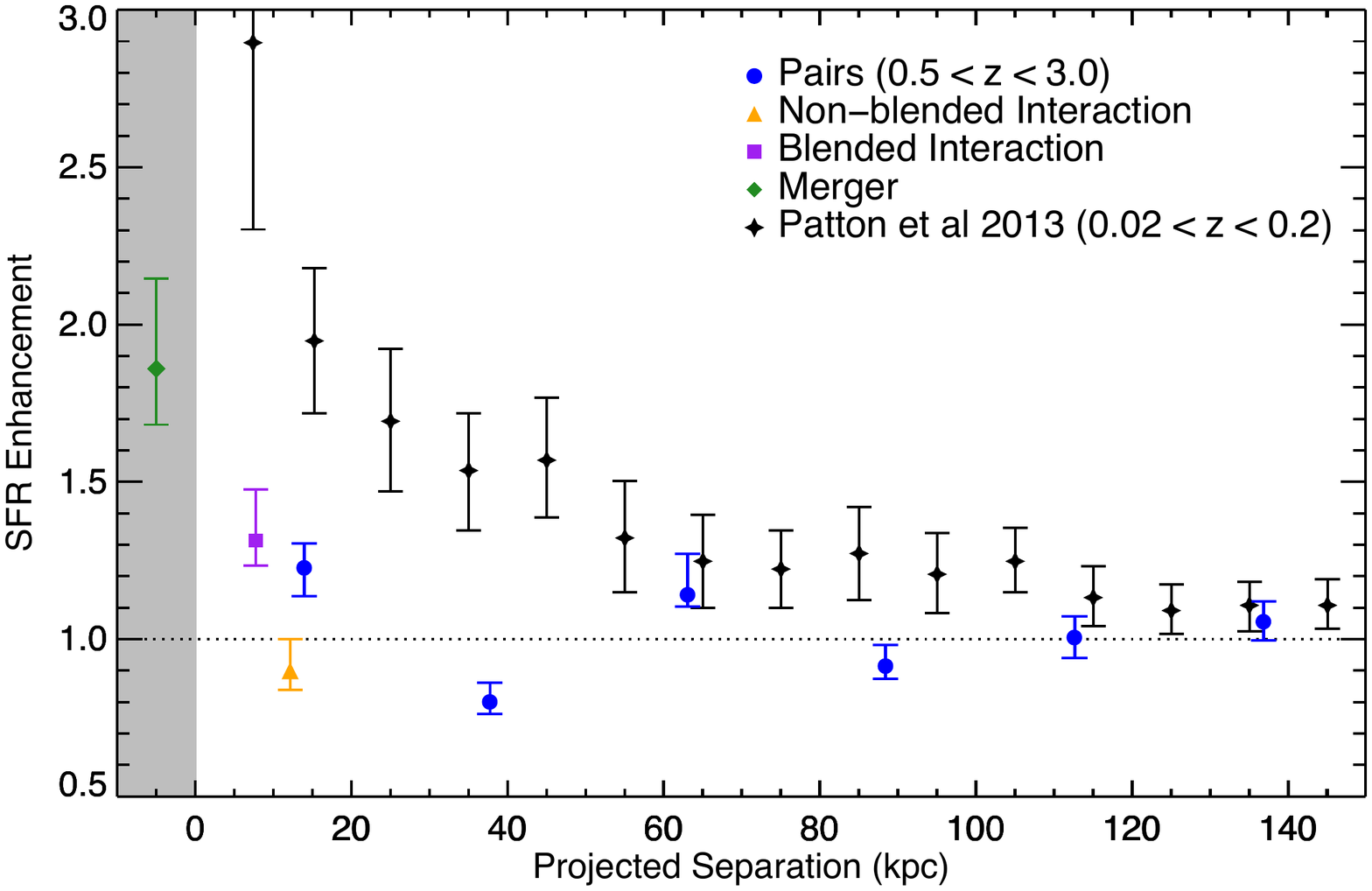}
    \caption{Comparison of the SFR enhancement as a function of the projected separation of interactions and mergers ($0.5<z<3.0$) to the pair sample in the local universe ($0.02<z<0.20$) from \citet{patton2013}. SFR enhancement is defined as the ratio of the average SFR of the pair sample to that of the control sample (Equation \ref{eq:enh}.) The dotted line corresponds to a value of one, which represents no interaction-induced SFR enhancement. The blue filled circles show the SFR enhancement for our complete galaxy pair sample ($log(M_{pair}/M_\odot)>10.3$ and $0.5<z<3.0$). The filled green diamond, filled purple square, and filled orange triangle correspond to the SFR enhancements for our visually identified merger sample, blended interaction sample, and non-blended interaction sample, respectively. The gray region corresponds to the merger (or post merger) stage. The filled black stars correspond to the \citet{patton2013} SFR enhancement results for a pair sample in the local ($0.02<z<0.20$) universe selected from SDSS observations. While we see significant enhancement for galaxy mergers in our high redshift sample, the overall level of enhancement is less than that seen in the local pair sample, indicative of a change in role of interactions and mergers in inducing star formation over cosmic time.}
    \label{fig:sfr_enh_comp}
\end{figure*}

In contrast to our results at $0.5<z<3.0$, there are several studies in the local universe ($z\sim0$) that find considerable SFR enhancement in galaxy interactions by comparing the star formation of interacting and isolated (control) galaxies (e.g., \citet{alonso2004, woods2007,ellison2008, xu2010,ellison2013a, patton2013}). Almost all of these studies find the strongest enhancement for pairs with projected separation less than 30\thinspace kpc.

The closest study on SFR enhancement in the local universe to our high redshift study is \citet{patton2013} (P13 from here on). Therefore, we compare our SFR enhancement results with P13 results to understand how the level of SFR enhancement differs at high redshift compared to the local universe. P13 measure the SFR enhancement for local pairs ($0.02<z<0.2$ and stellar mass ratio $<10$) identified using SDSS observations. We compare their SFR enhancement results with our results in Figure~\ref{fig:sfr_enh_comp}. The highest level of SFR enhancement that they find is a factor of $\sim 2.9$ for pairs with projected separation less than 10\thinspace kpc. Their estimated SFR enhancement value for pairs with projected separation between 10\thinspace kpc and 20\thinspace kpc is $\sim 1.95$, which is a factor of 1.6 times ($\sim 3.2\sigma$) higher than our SFR enhancement result of 1.23$^{+0.08}_{-0.09}$ for pairs with projected separation between 0\thinspace kpc and 25\thinspace kpc (median $\sim14$\thinspace kpc). P13 see a clear trend of increasing SFR enhancement with decreasing projected separation. We find that this trend is absent in our for pairs with projected separations of $>$25 kpc based on Figure~\ref{fig:sfr_enh_comp}, at separations $<$25 kpc, we see an overall trend of increasing SFR enhancement with decreasing projected separation. This is particulary notable for our visually identified merger and interaction sample. For the SFR enhancement our blended interactions sample, we see a value slightly less than one, which could be due to low number statistics. Unlike our results, P13 observe an enhancement in SFR for pairs with projected separations of up to 150\thinspace kpc. Many factors can impact the measurement of SFR enhancement (such as redshift evolution, sample selection, SFR measurement method, etc.) that would be cumulatively responsible for the differences between our results and P13 results. 

Apart from the clear differences in the redshift range of our pair sample ($0.5<z<3.0$) and their pair sample ($0.02<z<0.2$), the P13 pair sample also consists of both major and minor pairs (mass ratio $<10$), unlike our sample of only major galaxy pairs. They also apply a stricter line of sight velocity difference constraint to identify galaxy pairs ($\Delta V<300$\thinspace km s$^{-1}$) compared to our relatively liberal constraint ($\Delta V<1000$\thinspace km s$^{-1}$). Furthermore, similar to many local studies, P13 select only star forming galaxies identified using emission lines, which means that their sample consists of only star-forming galaxies compared to our sample which also contains non-star forming galaxies. If we only consider the subset of our galaxies that are star-forming, our SFR enhancement result (Figure~\ref{fig:sfr_enh_sfdiv}) of 1.32$^{+0.09}_{-0.11}$ at 0-25\thinspace kpc is 1.5$\times$ (compared to 1.6$\times$ for the complete sample), lower than P13 SFR enhancement at 10-20\thinspace kpc. Hence, even for the star-forming sample, our pair sample ($0.5<z<3.0$) shows less SFR enhancement than a local ($0.02<z<2.0$) pair sample P13 at similar projected separations.

Similarly, \citet{ellison2013b} find a SFR enhancement of $\sim3.5$ in a post-merger sample compared to the controls generated using SDSS observations. Their SFR enhancement is $\sim1.9\times$ higher than our visually identified merger sample ($0.5<z<3.0$). For our sample divided into two redshift bins, the local enhancement for coalesced mergers is $1.4\times$ and $2.2\times$ higher than our $0.5<z<1.6$ and $1.6<z<3.0$ samples, respectively. This suggest evolution in the merger-induced star formation enhancement between these three redshift bins.

\subsection{Comparison with Simulations}
 
A relative decrease in the SFR enhancement level and its duration in merging galaxies with a significantly higher gas fraction has been suggested by some studies conducted using idealized binary merger simulations \citep{bournaud2011,hopkins2013,scudder2015,fensch2017}. For example, \citet{fensch2017} use idealized binary merger simulations to study the effect of a galaxy merger on the SFR of galaxies and show that the amount and the duration of the merger-induced star formation excess is about ten times lower for a high redshift galaxy (gas fraction $\sim$ 60\%, $z\sim2$) mergers compared to their low redshift counterparts (gas fraction $\sim10\%$, $z\sim0$). \citet{fensch2017} suggest that the high turbulence in these systems makes further compression of gas and generation and propagation of inflows weaker at high-$z$ than in low-$z$ interactions, causing a considerable decrease in the ability of mergers to increase the SFR of galaxies at high-$z$.

Similarly, using a suite of nine binary merger simulations with the same orbital parameters but with varying gas fractions ($0.04\le M_{*}/M_{gas}\le1.78$), \citet{scudder2015} find an anti-correlation between the SFR enhancement at coalescence and the pre-merger gas fraction of galaxies. However, such studies based on binary merger simulations usually do not contain cosmic gas inflows and have a very limited sample size. Therefore, we also compare our results with SFR enhancement studies based on hydrodynamical cosmological simulations. Many of these studies also show the weakening of enhancement levels with increasing redshift. For example, \citet{patton2020} present a study (similar to their observational study P13) of interacting galaxy pairs identified in the IllustrisTNG cosmological simulations. They select massive ($M_*>10^{10}M_{\odot}$) galaxy pairs at $0\leq z<1$ and their corresponding stellar mass-, redshift-, and environment-matched controls and see a gradual decrease in the sSFR enhancement levels with increasing redshift in the same projected separation bin.

\citet{martin2017} also find a decrease in merger-induced SFR enhancement level with increasing redshift in the Horizon-AGN cosmological simulations. This reduction in SFR enhancement is also suggested by our higher SFR enhancement result (2.40$^{+0.62}_{-0.37}$) for the visually identified merger sample at low $z$ ($0.5<z<1.6$) compared to a lower SFR enhancement (1.58$^{+0.29}_{-0.20}$) at high $z$ ($1.6<z<3.0$) shown Figure~\ref{fig:sfr_enh_ls_zdiv}. Comparisons of our results with the local results, as discussed earlier, also show this reduction in SFR enhancement with redshift. 

\citet{hani2020} also use IllustrisTNG (TNG300-1) to identify post-mergers at $0\leq z\leq1$ and a well-matched control sample. They see a SFR enhancement of a factor of $\sim2$ in their merger sample compared to the control sample, which is within error bars of our SFR enhancement result (2.40$^{+0.62}_{-0.37}$) for the visually identified merger sample at $z<1.6$ (left panel of Figure~\ref{fig:sfr_enh_ls_zdiv}). They match SF mergers with SF controls, which is analogous to our results for SF mergers (Figure~\ref{fig:sfr_enh_ls_only_sf_zdiv}), where our SFR enhancement (2.08$^{+0.42}_{-0.15}$) for the merger sample at low $z$ ($0.5<z<1.6$) is almost same as their SFR enhancement ($\sim2$).

While \citet{hani2020} see no redshift evolution in their SF mergers compared to SF controls over $0\leq z<1$, we see a $1.75\times$ higher SFR enhancement at low-$z$ ($z<1.6$) compared to high-$z$ ($z>1.6$). Hence, the comparison suggests that there might be a mild evolution of SFR enhancement with redshift, which becomes detectable when enhancement is compared over a larger redshift range ($0.5<z<3.0$ in our study), and which could be missed when enhancement is compared over a smaller redshift range \citep[$0\leq z\leq 1$ in][]{hani2020}. For a larger redshift range, our SFR enhancement value (1.86$^{+0.29}_{-0.18}$) for the visually identified merger sample at $0.5<z<3.0$ is within the error bars of the sSFR enhancement in a merger sample compared to a mass-matched control sample  at $0<z<2.5$ in the Simba Simulation \citep{rodrigues2019}.

\subsection{Comparison with Previous High Redshift Studies}

There are also observational studies exploring the effects of interactions and mergers on the SFR of galaxies at high redshift. For example, \citet{lackner2014} identify merging galaxies in the COSMOS field at $0.25<z<1.00$ with log $M_{*}/M_{\odot}>10.6$ by applying an automated method of median-filtering the high-resolution COSMOS HST images to distinguish two concentrated galaxy nuclei at small separations (2.2-8\thinspace kpc). This method is sensitive to very close pairs and advanced stage mergers with double nuclei and hence are all roughly at a similar merger stage. We note that because of the differences in the selection criteria, some objects in their sample could be included in our visually identified merger and blended interaction samples. They find a SFR enhancement of value of 2.1$\pm$0.6 in their merging sample compared to the non-merging sample. Their result is consistent with our result for the visually identified merger sample (2.40$^{+0.62}_{-0.37}$) and blended-interaction sample (1.44$^{+0.35}_{-0.12}$) at $0.5<z<1.6$ (left panel of Figure~\ref{fig:sfr_enh_ls_zdiv}). This comparison also shows the importance of comparing the definition or identification criteria (e.g., merging galaxies in this case) when comparing the results of two studies. Our SFR enhancement result for the visually identified merger sample at $z>1.6$ is also consistent with the sSFR enhancement ($\sim 2.2$) in mergers compared to non-interacting galaxies at ($z\sim2$) found in \citet{kaviraj2013}.

In contrast to these results, \citet{wilson2019} do not see a significant SFR enhancement in merging galaxy pairs compared to a control sample of isolated galaxies at $1.5\lesssim z\lesssim3.5$. The differences in our results could be due to differences in the pair selection and control matching criteria used. They identify pairs as two objects whose spectra were obtained on the same Keck/MOSFIRE slit. Their 30 galaxy pairs have projected separations less than 60\thinspace kpc, relative velocities less than 500\thinspace km s$^{-1}$, and stellar mass ratios ranging from 1.1 to 550 ($<3$ for 40\thinspace $\%$ of pairs). These criteria mean that their sample consists of major interactions (mass ratio $<4$), minor interactions (mass ratio between 4 and 10), and systems with even larger mass ratios; hence, a much larger difference between the stellar mass of the primary and secondary galaxies, which can significantly dilute the estimated interaction effects. 

Similarly, \citet{pearson2019} train and use convolutional neural networks to identify over 200,000 galaxies at $0.0<z<4.0$ in the SDSS, KiDS, and CANDELS surveys as merging or non-merging galaxies. They then compare the SFR of merging and non-merging galaxies and find a slight enhancement of $\sim1.2$ in merging galaxies compared to their non-merging sample. Our SFR enhancement results for pairs in the closest projected separation bin as well the blended-interactions in strong agreement with their SFR enhancement results. However, we also note a large difference in the redshift range and other selection criteria, making a direct comparison difficult. 

\subsection{Variation in SFR Enhancement with Stellar Mass}
Some studies also find a decrease in the SFR enhancement or sSFR enhancement with the stellar mass of galaxies. For example, \citet{hani2020} find a change in sSFR enhancement with stellar mass. For $M_*<10^{11}M_{\odot}$ they see an sSFR enhancement of 1.5-2.5 (depending on the TNG simulation used), which reduces to $\sim 1$ at higher masses. The same trend is also seen for mergers in the Simba Simulation \citep{rodrigues2019}. A similar trend of decreasing level of SFR enhancement with increasing stellar mass is also seen in the merger sample of \citet{silva2018}. Our SFR enhancement results for galaxy pairs with $M_*<10^{10.88}M_{\odot}$ (1.51$^{+0.11}_{-0.10}$) and $M_*>10^{10.88}M_{\odot}$ (0.96$^{+0.10}_{-0.14}$) shown in Figure~\ref{fig:sfr_enh_mdiv} suggest a similar trend as seen in the above-mentioned studies. 

As discussed in \citet{shah2020}, selecting a well-matched control sample is crucial for this type of analysis. While we have only selected isolated galaxies that do not show obvious visual signs of interactions or mergers as control candidates, our control sample can unintentionally still contain merging galaxies that were not identified as mergers in the visually identified sample. This is mainly due to the difficulty in identifying mergers and interactions at high redshift due to weak or non-detectable merger signatures. This potential contamination of mergers or interactions in the control sample would dilute the estimated SFR enhancement in our sample. Hence, the estimated SFR enhancement values are most likely lower limits of the actual SFR enhancements.

\section{Summary} \label{sec:obs_sfr_summary}

The goal of this study is to understand the effect of galaxy interactions and mergers on the star formation of galaxies at high redshift and compare the SFR enhancement results with the local interaction results in order to understand evolutionary effects on the role of interactions in driving star formation activity. We used our sample of major spectroscopic galaxy pairs, visually identified interactions and mergers, and their corresponding control samples at $0.5<z<3.0$, as described in detail in \citet{shah2020}. The samples were generated using deep multi-wavelength photometric and spectroscopic observations from the CANDELS and COSMOS surveys. Our 1327 (2351) spectroscopic galaxy pairs satisfy five criteria: a spectroscopic redshift in the range $0<z<3$, a relative line of sight velocity less than 1000 (5000) \thinspace km s$^{-1}$, stellar mass of each of the galaxies in a pair more than 10$^{10}M_{\odot}$, stellar mass ratio of the primary (more massive) to the secondary galaxy less than four, and projected separation less than 150 kpc. Our controls are closely matched to individual paired galaxies in their stellar mass, redshift, and environment density. We estimate the SFR enhancement in the galaxy pair sample by taking the ratio of the weighted mean of the SFR of the galaxy pair sample over that of the corresponding control sample. Our main findings are:

\begin{enumerate}
    \item We see evidence for a slight enhancement of a factor 1.23$^{+0.08}_{-0.09}$ ($\sim2.6\sigma$) in the closest projected separation bin ($d<25$\thinspace kpc) for our full galaxy pair sample (Figure~\ref{fig:sfr_enh_pairs}). However, we also see a significant amount of scatter around the SFR enhancement value of one at all separation bins. Therefore, in contrast to results for local pair studies, we do not see a clear trend of increasing SFR enhancement with decreasing projected separation in our spectroscopic galaxy pair sample.

    \item We divide our pair sample by redshift and find that at the closes separation ($<25\thinspace kpc$), the low-$z$ ($0.5<z<1$ sample has a higher SFR enhancement than the high-$z$ ($1<z<3.0$) sample by a factor of $\sim 1.3$ (2.2$\sigma$)  (Figure~\ref{fig:sfr_enh_zdiv}). Though this difference is marginally significant, it is consistent with interaction-induced star formation having a decreased role at higher redshift.
        
    \item We see an enhancement level of a factor of 1.86$^{+0.29}_{-0.18}$ ($\sim3\sigma$) in our visually identified merger sample. There is a clear trend of increasing SFR enhancement with decreasing projected separation in our visually identified sample of interactions and mergers as shown in Figure~\ref{fig:sfr_enh_ls}. The sample shows this trend at both low-$z$ ($0.5<z<1.6$) and high-$z$ ($1.6<z<3.0$) (2.40$^{+0.62}_{-0.37}$ vs. 1.58$^{+0.29}_{-0.20}$; Figure~\ref{fig:sfr_enh_ls_zdiv}), with the level of enhancement decreasing with increasing redshift. This result again suggests evolution in the level of merger-driven star formation with time.
        
    \item We see an enhancement of 1.51$^{+0.11}_{-0.10}$ ($\sim4.6\sigma$) for our pairs with a lower mass primary galaxy ($M_{prim}<10.88\,M_{\odot}$) compared to the enhancement (0.96$^{+0.10}_{-0.14}$) in  pairs with a higher mass primary galaxy ($M_{prim}>10.88\,M_{\odot}$) in the same closest projected separation bin. This factor of 1.6 difference in the enhancement level hints at stronger effects of interactions in enhancing SF of lower mass galaxy pairs. This is consistent with the results of both observational \citep[e.g.,][]{silva2018} and simulation  \citep[e.g.,][]{rodrigues2019} studies. 
    
    \item We also see a slightly higher level of SFR enhancement (1.32$^{+0.09}_{-0.11}$ vs.\ 0.84$^{+0.12}_{-0.10}$) in the star-forming paired galaxy sample compared to the non-star forming paired galaxy sample in the closest projected separation bin. For blended pairs, we used the combined SFR of the system to check if it is star-forming or non-star forming. 
   
\end{enumerate}

Overall, our results show a slight SFR enhancement in close pairs and a significant enhancement in advanced stage mergers at $0.5<z<3.0$, and an absence of SFR enhancement at larger pair separations ($>25$\thinspace kpc). Comparison of our results with local studies (Figure~\ref{fig:sfr_enh_comp}) suggests that the effect of interactions and mergers on SFR weakens at high $z$, which is consistent with the predictions of some simulations. Our study on SFR enhancement in a large sample of spectroscopic galaxy pairs and mergers provides a deeper understanding of the role of galaxy mergers and interactions in galaxy evolution in the high redshift universe. Future spectroscopic surveys of close pairs at high redshift are required, especially at $z>2$, to increase the sample size and improve statistics to enable a more comprehensive analysis of the evolving role of galaxy interactions over cosmic time.

\section*{Acknowledgments}

Support for this work was provided by NASA through grants HST-GO-13657.010-A and HST-AR-14298.004-A awarded by the Space Telescope Science Institute, which is operated by the Association of Universities for Research in Astronomy, Inc., under NASA contract NAS 5-26555. Support was also provided by NASA through grant NNX16AB36G as part of the Astrophysics Data Analysis Program. This work was also supported by start-up funds and the Dean's Research Initiation Grant fund from the Rochester Institute of Technology's College of Science. Spectral energy distribution fitting was performed using the computational resources and support from Research Computing Services at the Rochester Institute of Technology \citep{https://doi.org/10.34788/0s3g-qd15}. ES thanks the LSSTC Data Science Fellowship Program, which is funded by LSSTC, NSF Cybertraining Grant \#1829740, the Brinson Foundation, and the Moore Foundation; The participation of ES in the program has benefited this work. DRP acknowledges financial support from NSERC of Canada. ET acknowledges support from CATA-Basal AFB170002 and FB210003, FONDECYT Regular grant 1190818, and Millennium Nucleus NCN19\_058 (TITANs). BL acknowledges support from the National Aeronautics and Space Administration under NASA grant no. 80NSSC21K0986. This paper does not reflect the views or opinions of the National Science Foundation or the American Association for the Advancement of Science (AAAS).

This work was supported by a NASA Keck PI Data Award, administered by the NASA Exoplanet Science Institute. Some of the data presented herein were obtained at the W. M. Keck Observatory, which is operated as a scientific partnership among the California Institute of Technology, the University of California and the National Aeronautics and Space Administration. The Observatory was made possible by the generous financial support of the W. M. Keck Foundation. 

Based in part on observations obtained at the international Gemini Observatory and processed using the Gemini IRAF package, a program of NOIRLab, which is managed by the Association of Universities for Research in Astronomy (AURA) under a cooperative agreement with the National Science Foundation. on behalf of the Gemini Observatory partnership: the National Science Foundation (United States), National Research Council (Canada), Agencia Nacional de Investigaci\'{o}n y Desarrollo (Chile), Ministerio de Ciencia, Tecnolog\'{i}a e Innovaci\'{o}n (Argentina), Minist\'{e}rio da Ci\^{e}ncia, Tecnologia, Inova\c{c}\~{o}es e Comunica\c{c}\~{o}es (Brazil), and Korea Astronomy and Space Science Institute (Republic of Korea). The authors wish to recognize and acknowledge the very significant cultural role and reverence that the summit of Maunakea has always had within the indigenous Hawaiian community.  We are most fortunate to have the opportunity to conduct observations from this mountain.

Based in part on observations made with the NASA/ESA Hubble Space Telescope, obtained from the Data Archive at the Space Telescope Science Institute, which is operated by the Association of Universities for Research in Astronomy, Inc., under NASA contract NAS 5-26555. This work is based in part on observations made with the Spitzer Space Telescope, which is operated by the Jet Propulsion Laboratory, California Institute of Technology under a contract with NASA. 

\bibliography{references.bib}
\end{document}